# Random taste heterogeneity in discrete choice models: Flexible nonparametric finite mixture distributions

Akshay Vij (corresponding author)
Institute for Choice
University of South Australia
Level 13, 140 Arthur Street
North Sydney, NSW 2060, Australia
vij.akshay@gmail.com

Rico Krueger
Research Centre for Integrated Transport Innovation
School of Civil and Environmental Engineering
University of New South Wales
Sydney, NSW 2052, Australia
r.krueger@student.unsw.edu.au

**Abstract**

This study proposes a mixed logit model with multivariate nonparametric finite mixture distributions. The support of the distribution is specified as a high-dimensional grid over the coefficient space, with equal or unequal intervals between successive points along the same dimension; the location of each point on the grid and the probability mass at that point are model parameters that need to be estimated. The framework does not require the analyst to specify the shape of the distribution prior to model estimation, but can approximate any multivariate probability distribution function to any arbitrary degree of accuracy. The grid with unequal intervals, in particular, offers greater flexibility than existing multivariate nonparametric specifications, while requiring the estimation of a small number of additional parameters. An expectation maximization algorithm is developed for the estimation of these models. Multiple synthetic datasets and a case study on travel mode choice behavior are used to demonstrate the value of the model framework and estimation algorithm. Compared to extant models that incorporate random taste heterogeneity through continuous mixture distributions, the proposed model provides better out-of-sample predictive ability. Findings reveal significant differences in willingness to pay measures between the proposed model and extant specifications. The case study further demonstrates the ability of the proposed model to endogenously recover patterns of attribute non-attendance and choice set formation.

# 1. Introduction

The field of discrete choice analysis has long wrestled with the question of how best to represent heterogeneity in the decision-making process (for a recent review of the literature, the reader is referred to Yuan et al., 2015). In cases where tastes vary systematically with observable variables, heterogeneity may be captured through interactions between observable characteristics of the decision-maker and observable attributes of the alternatives. However, capturing heterogeneity systematically may be insufficient when tastes vary with unobservable variables or purely randomly, and can result in inconsistent parameter estimates (Chamberlain, 1980). In such cases, heterogeneity in the decision-making process may be captured through additional interactions between observable variables and the stochastic component.

Mixed logit is by far the most popular discrete choice model for incorporating random taste heterogeneity. The model specifies choice probabilities as a mixture of multinomial logit probabilities. McFadden and Train (2000) show that mixed logit can approximate choice probabilities given by any discrete choice model derived from the theory of random utility maximization to any desired level of closeness. However, the "theorem is an existence proof only and does not provide guidance for finding the mixing distribution that attains an arbitrarily close approximation" (Train, 2008).

Most mixture distributions employed in practice can broadly be classified as one of two types: parametric or nonparametric. Parametric distributions have clearly defined functional forms with a fixed number of parameters, as in the case of a normal or lognormal distribution. Though parametric distributions often provide an excellent fit to the data, they are limited by their functional form in the shapes that they can assume. There is no prior best distribution; studies usually estimate models with different distributions, and the most appropriate distribution is determined through a comparison across goodness of fit measures and behavioral interpretation. Due to the many different distributions that can possibly be estimated, the process can be labor-intensive, and any search for the most appropriate distribution is necessarily ad hoc in that "not every possible distribution can be considered" (Keane and Wasi, 2013).

Nonparametric distributions do not have well-defined functional forms with a fixed number of parameters. Rather, the number of parameters increases and the functional form grows in complexity as more data become available, as in the case of latent class choice models (LCCMs). Seminonparametric distributions are closely related to nonparametric distributions, and may be thought of as nonparametric mixtures of parametric distributions, such as finite mixtures of normal distributions (used, for example, by Train, 2008 and Fosgerau and Hess, 2009). Nonparametric and seminonparametric distributions can asymptotically mimic any multivariate distribution, but computational constraints generally preclude the estimation of models with a high degree of complexity, and models estimated in practice tend to be simpler abstractions that are unable to leverage the flexibility offered by the general framework.

The objective of this study is to develop a fully flexible and computationally tractable way of incorporating random taste heterogeneity within existing discrete choice models. We propose using a mixed logit model with a nonparametric mixture distribution, where the support of the distribution is specified as a high-dimensional grid over the coefficient space, with equal or unequal intervals between successive points along the same dimension, and the location of each point on the grid and the probability mass at that point are model parameters that need to be estimated. First proposed by Dong and Koppelman (2014), the framework does not require the analyst to specify the shape of the distribution prior to model estimation, and the specification can approximate any multivariate probability distribution function to any arbitrary degree of accuracy. However, empirical applications in the literature have thus far been limited to distributions over low-dimensional coefficient spaces with a small number of mass points, likely due to the computational burden imposed by standard gradient-based maximum likelihood estimation routines employed by the study, and the value of the framework over other parametric and nonparametric distributions is unclear. We outline an expectation maximization (EM) algorithm for the estimation of the proposed model that is able to estimate behaviorally meaningful specifications over high-dimensional coefficient spaces with hundreds of mass points. The performance of the estimation algorithm is tested using multiple synthetic datasets, and the benefits of the model specification are evaluated using a case study on travel mode choice behavior.

Despite significant recent advances in discrete choice methods, the question of how best to incorporate random taste heterogeneity has remained an open line of enquiry. Since 2010 alone, we are aware of at least eleven new methods that have been proposed in the literature (c.f. Bastin et al., 2010; Bujosa et al., 2010; Fiebig et al., 2010; Bhat and Sidharthan, 2012; Bastani and Weeks, 2013; Greene and Hensher, 2013; Keane and Wasi, 2013; Dong and Koppelman, 2014; Train, 2016; Bansal et al., 2017; Bhat and Lavieri, 2017). However, for reasons expanded upon in the following section, each of these methods has proven inadequate in one way or another. This study contributes to the literature through the development of a model framework and estimation algorithm that can help overcome some of the constraints arising from these and other specifications. In particular, compared to extant multivariate nonparametric specifications, we show how our proposed approach of specifying the parameter space as a high-dimensional grid with unequal intervals can offer considerable improvements in flexibility, while requiring the estimation of a small number of additional parameters.

The remainder of the paper is structured as follows: Section 2 reviews the relevant literature on the use of mixture distributions for discrete choice analysis; Section 3 presents the methodological framework; Section 4 describes how the framework may be estimated in practice using the EM algorithm; Section 5 details findings from three Monte Carlo experiments evaluating the ability of the proposed framework to recover known parametric univariate and bivariate distributions, respectively; Section 6 applies the framework to a study on travel mode choice behavior, and compares findings from the framework with extant approaches for incorporating random taste heterogeneity; Section 7 examines the robustness of the estimation method to different starting values for the model parameters; and Section 8 concludes the paper with a summary of key results and directions for future research.

## 2. Literature Review

Section 2.1 surveys studies that have used parametric mixture distributions and Section 2.2 surveys studies that have used nonparametric and seminonparametric mixture distributions. Throughout, we identify advantages and disadvantages to each of the many different mixture distributions that have been used in the literature, and how they motivate the particular framework employed by this study.

### 2.1 Parametric Mixture Distributions

Early applications of the mixed logit model to incorporate random taste heterogeneity employed commonly known parametric distributions, such as the normal and lognormal, and limited their attention to one or two randomly distributed taste coefficients (see, for example, Ben-Akiva et al., 1993). Advances in computational power and corresponding leaps in simulation methods have since helped set off a veritable explosion in the development and application of these models (see Train, 2009). Numerous functional forms have been used in the literature, such as the censored normal, triangular and $S_B$ distributions (see, for example, Train and Sonnier, 2005 and Keane and Wasi, 2013), and attempts have also been made to describe these distributions as functions of covariates to improve fit and ease interpretation (see, for example, Harris and Keane, 1999; Bhat, 2000; and Greene et al., 2006).

While the use of parametric distributions often provides an excellent fit to the data (see, for example, Allenby and Rossi, 1998; Andrews et al., 2002; and Greene and Hensher, 2003), there are three major limitations. First, parametric distributions are limited by their functional form in the shapes that they can assume. For example, the normal distribution, perhaps the most commonly employed distribution, is symmetric around the mean and has support on both sides of zero, rendering it inappropriate for coefficients that are asymmetric, signed or multimodal; skew normal distributions are asymmetric, but they too are unimodal and have support on both sides of zero; lognormal distributions are signed, but have thick tails; etc. Some studies have sought to overcome these limitations by using more esoteric distributions, such as truncated normal and $S_B$ distributions (Train and Sonnier, 2005), or combining multiple distributions, as in the case of the generalized multinomial logit model (Fiebig et al., 2010 and Hess and Rose, 2012). However, the original point stands in that the shapes that these more flexible distributions can assume is still limited.

Second, parametric distributions require the analyst to make a prior assumption about the mixture distribution for each randomly distributed coefficient. A wrongly specified distribution can have deleterious effects on coefficient estimates and the attendant model interpretation (c.f. Wedel et al., 1999 and Fosgerau, 2006). Since distributional assumptions exert influences of their own on the results (Hess and Rose, 2007), it has been argued that simply knowing that a coefficient is distributed randomly across respondents might be of limited utility to policy makers (Hess et al., 2009).

And third, parametric distributions estimated in the literature are usually univariate distributions that do not allow for more complex covariance structures. Multivariate parametric distributions with fully specified covariance matrices are significantly harder to estimate. For example, Bayesian procedures, such as the Markov chain Monte Carlo (MCMC) method, work well with multivariate normal distributions and their unbounded transformations, but the estimation of bounded distributions can prove considerably more difficult (Train and Sonnier, 2005). Similarly, the maximum approximate composite marginal likelihood (MACML) estimation approach proposed by Bhat (2011) and Bhat and Sidharthan (2012) allows for the efficient estimation of multivariate normal and skew-normal distributions, respectively, but neither study offers insights on how the approach may be extended to other distributions. Bhat and Lavieri (2017) combine the MACML estimator with traditional maximum simulated likelihood methods to estimate more flexible multivariate distributions, but the approach is subject to the same limitations as traditional maximum simulated likelihood estimation.

Multivariate parametric distributions that are easier to estimate often require the analyst to make a number of limiting assumptions about the covariance matrix. For example, the generalized multinomial logit (G-MNL) model specifies each coefficient as the product of two random variables with independent parametric probability density functions, such that the first random variable is specific to the coefficient and the second random variable is common to all coefficients (Fiebig et al., 2010 and Greene and Hensher, 2010). Though the resulting distribution allows for correlation between different pairs of coefficients (Hess and Rose, 2012), and the model can be reliably estimated using classical procedures, the unusual nature of the specification imposes constraints on the covariance matrix that may not necessarily hold true (for a comprehensive discussion on the limitations of the G-MNL model, refer to Hess and Train, 2017).

## 2.2 Nonparametric and Seminonparametric Distributions

Efforts to relax some of the restrictions arising from the use of parametric mixture distributions have led to interest in nonparametric and seminonparametric distributions. Nonparametric and seminonparametric distributions do not require the analyst to make prior assumptions about the shape of the distribution and can asymptotically mimic whatever shape best describes the heterogeneity in the data. However, the use of these distributions necessitates the estimation of a far larger number of parameters than parametric distributions, and the greater computational costs imposed by these distributions has proven to be a stumbling block to their widespread adoption. Sections 2.2.1 and 2.2.2 describe discrete choice studies that have employed nonparametric and seminonparametric distributions, respectively.

### 2.2.1 Nonparametric Distributions

The most popular nonparametric distribution is perhaps the finite mixture distribution employed by LCCMs, where the support of the distribution is defined as a fixed number of points in a high-dimensional coefficient space, and the location of each point and the probability mass at that point are model parameters to be estimated (for a comprehensive reference on finite mixture models in general, the reader is referred to McLachlan and Peel, 2004). LCCMs were first developed in the field of marketing sciences as tools to identify relatively homogenous consumer segments that differ substantially from each other in terms of their behavior in the marketplace (Kamakura and Russell, 1989), and have since found widespread popularity in other applied disciplines. In general, the more mass points, or classes, that an LCCM has, the more flexible will be the resulting mixture distribution. However, the estimation of LCCMs with a high number of classes can be computationally difficult (Yuan et al., 2015), and most LCCMs estimated in practice usually restrict themselves to three or four classes, with thirty classes being the most that the author is aware of (Train, 2008). It has been argued that LCCMs "with a small number of mass points may inadequately capture the full extent of heterogeneity in the data" (Allenby and Rossi, 1998).

An alternative formulation tested by Train (2008), based on a framework first proposed by Bajari et al. (2007), uses a finite mixture distribution where the support of the distribution is fixed by the analyst prior to estimation, usually as a high-dimensional grid in the coefficient space, and the probability masses at each support point are the only model parameters that need to be estimated. Results from a case study on alternative-fueled vehicle choice behavior are promising in that restricting the location of the mass points is found to allow for the efficient estimation of distributions with significantly more mass points than would otherwise be possible. For example, Train (2008) is able to estimate distributions with as many as 233,280 mass points in as little as 31 minutes, providing greater flexibility in the shape of the distribution than any of the other methods used previously in the literature. However, a major limitation to the approach is that model performance, as measured by both goodness of fit and behavioral interpretation, varies considerably depending upon the predetermined location of the mass points. Possible solutions proposed by Train (2008), Bastani and Weeks (2013) and Train (2016) all require the determination of the mass point locations exogenously, through prior information possibly gathered from the estimation of simpler but more restrictive frameworks.

A natural and more feasible alternative would be to estimate the location for each of the mass points endogenously, along with the probability mass at each point. Dong and Koppelman (2014) propose such a model specification, where the support of the distribution is specified as a high-dimensional grid over the coefficient space, and the location of each point on the grid and the probability mass at that point are model parameters that need to be estimated. However, the framework is only evaluated for a two-dimensional coefficient space, due possibly to the computational intractability of standard gradient-based maximum likelihood estimation routines employed by the study. As a result, the value of the framework over a traditional LCCM is not readily apparent. We attempt to overcome this limitation through the development of an EM algorithm for the estimation of the same model. In subsequent sections, we demonstrate the algorithm's ability to feasibly estimate behaviorally meaningful models with hundreds of mass points over high-dimensional coefficient spaces, and we illustrate the benefits of the model framework and estimation routine through a case study on travel mode choice behavior.

### 2.2.2 Seminonparametric Distributions

For the sake of completeness, we also review mixed logit models that have used seminonparametric distributions. The most common such model is the mixture of distributions model, also referred to in the literature as latent class random parameters logit model and latent class mixed multinomial logit model. Like nonparametric distributions, finite mixtures of parametric distributions too can asymptotically mimic any shape, but high computational costs often preclude the estimation of models with a large number of mixtures, and models estimated in practice are typically less flexible than those with nonparametric distributions. For example, Fosgerau and Hess (2009) and Bujosa et al. (2010) use discrete mixtures of normal distributions, and Greene and Hensher (2013) use discrete mixtures of triangular distributions, but each of these studies is limited by its attention solely to univariate distributions. Train (2008) expands the framework to allow for mixtures of multivariate normal distributions and their transformations, such as multivariate lognormal and truncated normal distributions, but the framework is empirically evaluated only for the case of mixtures of two independent multivariate distributions.

The mixture of distributions model belongs to the broader family of sieve estimators, "defined as an estimator which approximates unknown functions with a series of basis functions" (Yuan et al., 2015). In the case of the mixture of distributions model, "the basis functions are the base distribution of the mixture, such as the normal in a mixture of normals." Other sieve estimators used in the literature include series approximations with Legendre polynomials (Fosgerau and Bierlaire, 2007) and B-spline approximations with cubic polynomials (Bastin et al., 2010). However, like the mixture of distributions model, these other approximations too suffer from restrictions. For example, the two studies cited here both only consider univariate distributions, and the feasibility of multivariate polynomial approximations is left as a subject for future research. Fosgerau and Mabit (2013) propose a polynomial series approximation to multivariate distributions, but the framework is not empirically evaluated for the multivariate case.

Train (2016) proposes the logit-mixed logit model, an approximate generalization of seminonparametric and nonparametric distributions, as a way to feasibly estimate multivariate distributions. The mixing distribution has a finite and discrete support, like nonparametric distributions described in 2.2.1, but the probability mass at each point is a parametric function of the location of the mass point. Bansal et al. (2017) extend the framework to allow for a subset of the taste coefficients to be fixed parameters. While the proposed framework is able to easily recover complex multivariate distributional shapes, it requires the analyst to specify the range of each taste coefficient and the form of the probability mass function prior to model estimation.

## 3. Methodological Framework

Discrete choice models with nonparametric finite mixture distributions represent variations on the general LCCM framework, which can itself be viewed as part of the broader family of mixed logit models. In describing the proposed methodological framework, we begin with a description of the general mixed logit model. We present parametric continuous mixtures of logit models and nonparametric finite mixtures of logit models as special cases. We introduce the two nonparametric finite mixing distributions developed by this study, and we discuss their relationship with nonparametric mixing distributions employed by existing LCCMs.

The mixed logit model specifies the utility $u_{ntj}$ that decision-maker n derives from alternative j over observation t as follows:

$$u_{ntj|s} = \boldsymbol{\beta}_n^T \mathbf{x}_{ntj} + \varepsilon_{ntj} \tag{1}$$

where $\mathbf{x}_{ntj}$ is a $(K \times 1)$ vector of attributes of alternative j over observation t for decision-maker n; $\boldsymbol{\beta}_n$ is a $(K \times 1)$ vector of decision-maker n's taste coefficients; and $\varepsilon_{ntj}$ is the stochastic component of the utility specification, assumed to be i.i.d. Gumbel across decision-makers, observations and alternatives with location zero and scale one. In terms of notation, unbolded lower-case letters denote scalars, bolded lower-case letters denote vectors, and bolded upper-case letter denote matrices and sets.

Assuming that decision-makers are utility maximizers, the probability that decision-maker n chooses alternative j over observation t, conditional on $\boldsymbol{\beta}_n$, is given by the familiar logit expression:

$$P(y_{ntj} = 1 | \boldsymbol{\beta}_n) = \frac{\exp(\boldsymbol{\beta}_n^T \mathbf{x}_{ntj})}{\sum_{j' \in \mathbf{C}_{nt}} \exp(\boldsymbol{\beta}_n^T \mathbf{x}_{ntj'})} \tag{2}$$

where $y_{ntj}$ equals one if decision-maker n over observation t chose alternative j, and zero otherwise. Equation (2) may be combined iteratively over alternatives and observations to yield the following conditional probability of observing the vector of choices $\mathbf{y}_n$ for decision-maker n:

$$f_y(\mathbf{y}_n | \boldsymbol{\beta}_n) = \prod_{t=1}^{T_n} \prod_{j \in \mathbf{C}_{nt}} [P(y_{ntj} = 1 | \boldsymbol{\beta}_n)]^{y_{ntj}} \tag{3}$$

where $T_n$ denotes the number of observations for decision-maker n; and $\mathbf{C}_{nt}$ is the choice set for observation t and decision-maker n. In cases where the population distribution of $\boldsymbol{\beta}_n$ is represented by a parametric probability density function, denoted $f_\beta(\boldsymbol{\beta}_n | \boldsymbol{\theta})$ with parameter vector $\boldsymbol{\theta}$, the marginal probability may be derived as follows:

$$f_y(\mathbf{y}_n | \boldsymbol{\theta}) = \int f_y(\mathbf{y}_n | \boldsymbol{\beta}_n) f_\beta(\boldsymbol{\beta}_n | \boldsymbol{\theta}) d\boldsymbol{\beta}_n \tag{4}$$

For a given functional form $f_\beta(\boldsymbol{\beta}_n | \boldsymbol{\theta})$, estimates for the unknown model parameters $\boldsymbol{\theta}$ may be obtained by maximizing the likelihood function. And the appropriate functional form itself may be determined by comparing estimation results with different functional forms in terms of statistical measures of fit and behavioral interpretation.

In cases where the population distribution of $\boldsymbol{\beta_n}$ is represented by a nonparametric probability mass function defined over a finite support, the marginal probability may be given by the following discrete approximation to the parametric mixture model:

$$f_y(y_n) = \sum_{s=1}^{S} f_y(y_n|\boldsymbol{\beta_{ns}}) f_\beta(\boldsymbol{\beta_{ns}}) \tag{5}$$

where the support is defined as comprising up to S points, or classes, located in the K-dimensional coefficient space; and $f_\beta(\boldsymbol{\beta_{ns}})$ denotes the probability mass at the $s^{th}$ point for the $n^{th}$ decision-maker. The reader should note that equation (5) corresponds to the general LCCM framework, where $f_y(y_n|\boldsymbol{\beta})$ and $f_\beta(\boldsymbol{\beta_{ns}})$ denote the class-specific choice model and the class membership model, respectively. For a given number of classes, estimates for the location of the classes in the K-dimensional coefficient space and the class membership probabilities may be obtained by maximizing the likelihood function. And the appropriate number of classes itself may be determined by comparing estimation results with different numbers of classes in terms of statistical measures of fit and behavioral interpretation.

For the purposes of the proposed nonparametric finite mixture distributions, we assume that class membership probabilities are population parameters that are invariant across decision-makers:

$$f_\beta(\boldsymbol{\beta_{ns}}) = P(\boldsymbol{\beta} = \boldsymbol{\beta_{ns}}) = P(q_{ns} = 1) = \gamma_s \tag{6}$$

where we introduce the latent variable $q_{ns}$, such that it equals one if decision-maker n belongs to class s, and zero otherwise; $\boldsymbol{\gamma}$ is an $(S \times 1)$ vector of model parameters, the $s^{th}$ element of which, $\gamma_s$, denotes the probability associated with the $s^{th}$ class. The assumption is similar to most parametric mixture models estimated in practice, where the probability density $f_\beta(\boldsymbol{\beta_n}|\boldsymbol{\theta})$ is assumed to be the same across decision-makers.

Let **B** be the $(K \times S)$ matrix of taste coefficients, the $s^{th}$ column of which, $\boldsymbol{\beta_s}$, denotes the set of class-specific taste coefficients. We suppress the subscript n corresponding to the decision-maker, assuming implicitly that **B** is the same across decision-makers. Note that this does not imply that the model cannot capture patterns of systematic taste heterogeneity. The attribute vector $\mathbf{x_{ntj}}$ may include interactions between alternative and decision-maker characteristics, and the corresponding elements of **B** can capture sensitivities to the same. The case study in Section 6 employs just such a utility specification.

There are multiple ways in which **B** can be specified. We will be limiting our attention to three configurations, illustrated in Figure 1 for the case of a two-dimensional coefficient space. The first configuration employs an unstructured support, as illustrated by the two-dimensional example of Figure 1(A), where each element of **B** is an independent model parameter that needs to be estimated. It is the most straightforward approach, and the approach usually employed in the literature.

The second configuration specifies **B** as a grid with S points on the K-dimensional coefficient space, where distances between successive points along the same dimension are assumed to be equal, but interval lengths are allowed to differ across dimensions, as shown by the example of Figure 1(A). The vector of class-specific coefficients $\boldsymbol{\beta_s}$ for any class s is given by the location of the $s^{th}$ point on the grid, and can be described purely in terms of the location of one of the corners of the resulting K-orthotope, or hyperrectangle, denoted $\boldsymbol{\alpha}$, and the K-dimensional vector of maximal distances along each of the K dimensions, denoted $\boldsymbol{\delta}$:

$$\boldsymbol{\beta}_s = \boldsymbol{\alpha} + \mathbf{H}_s \boldsymbol{\delta} \tag{7}$$

where $\mathbf{H}_s$ is a (K × K) diagonal matrix of loadings. Let $\mathbf{H} = \{\mathbf{H}_1, \ldots, \mathbf{H}_S\}$. Note that $\mathbf{H}$ is not a model parameter that needs to be estimated, but is a function of the geometry that results from parameterizing the support of the probability distribution as a grid with equal intervals. For example, for the two-dimensional coefficient space shown in Figure 1(B), where the support is specified as a rectangular grid with 15 points, $\mathbf{H}_s$ is a (2 × 2) diagonal matrix for all s. Let class 8 denote the mass point in the third column from the left and the second row from the bottom, such that $\boldsymbol{\beta}_8 = (\alpha_1 + 2\delta_1/4, \alpha_2 + \delta_2/2)$. Then, the diagonal elements of $\mathbf{H}_8$ are both 0.5. The elements of each of the other matrices can be specified similarly by the analyst prior to estimation, and $\boldsymbol{\alpha}$, $\boldsymbol{\delta}$ and $\boldsymbol{\gamma}$ are the only model parameters to be estimated.

The third configuration also specifies $\mathbf{B}$ as a grid with S points on the K-dimensional coefficient space, but the distances between successive points along the same dimension are allowed to be unequal, as illustrated by the two-dimensional example of Figure 1(C). Let $\boldsymbol{\lambda}_k$ be an ($M_k$ × 1) vector whose elements denote the support of the marginal distribution along the $k^{th}$ dimension. In this case, the $k^{th}$ element of the vector of class-specific coefficients $\boldsymbol{\beta}_s$, denoted $\beta_{sk}$, can be described directly in terms of $\boldsymbol{\lambda}_k$:

$$\beta_{sk} = \mathbf{h}_{sk}^T \boldsymbol{\lambda}_k \tag{8}$$

where $\mathbf{h}_{sk}$ is an ($M_k$ × 1) vector of loadings. Let $\mathbf{h}_s = \{\mathbf{h}_{s1}, \ldots, \mathbf{h}_{sK}\}$ and $\mathbf{h} = \{\mathbf{h}_1, \ldots, \mathbf{h}_S\}$. Similarly, let $\boldsymbol{\lambda} = \{\boldsymbol{\lambda}_1, \ldots, \boldsymbol{\lambda}_K\}$. Note that $\mathbf{h}$ is not a model parameter that needs to be estimated, but is a function of the geometry that results from parameterizing the support of the probability distribution as a grid with unequal intervals. For example, for the two-dimensional coefficient space shown in Figure 1(C), where the support is specified as a rectangular grid with 12 points, $\mathbf{h}_{s1}$ and $\mathbf{h}_{s2}$ are (4 × 1) and (3 × 1) vectors for all s. Let class 7 denote the mass point in the third column from the left and second row from the bottom, such that $\boldsymbol{\beta}_7 = (\lambda_{13}, \lambda_{22})$. Then, $\mathbf{h}_{71} = \langle 0,0,1,0 \rangle$ and $\mathbf{h}_{72} = \langle 0,1,0 \rangle$. Each of the pairs of vectors for the other classes can be specified similarly by the analyst prior to estimation, and $\boldsymbol{\lambda}$ and $\boldsymbol{\gamma}$ are the model parameters to be estimated.

**Figure 1:** An example illustrating three different ways in which the support for a multivariate nonparametric finite mixture distribution may be specified for a two-dimensional coefficient space, and the minimum number of model parameters needed to describe the support of the resulting distribution

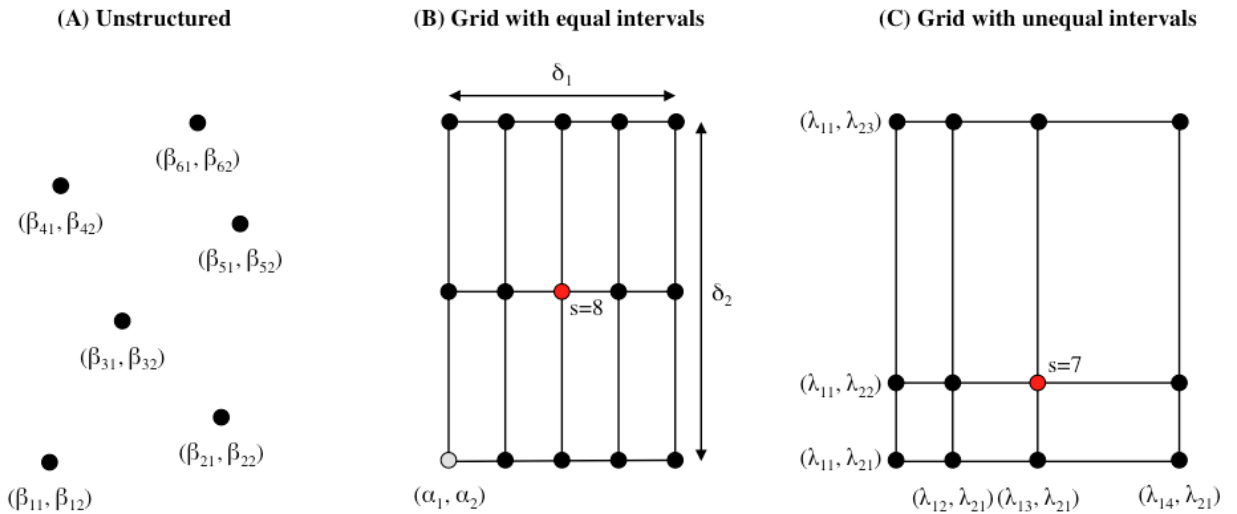

The second and third configurations may be thought of as casting a net in the coefficient space, where the net can expand or contract in size and move along the coefficient space as a function of the data, but the fineness of the net, as determined by the number of classes, is constrained prior to estimation. The three configurations are asymptotically equivalent. As the number of classes increases, either configuration can theoretically mimic any desired probability distribution function, and will asymptotically converge to the true underlying population distribution (Fox et al., 2012). In practice, as subsequent sections will demonstrate, it's often much easier to estimate models with a high number of classes using the second or the third specification than it is using the first. As mentioned previously, past studies that have estimated LCCMs with unstructured supports have been limited to models with three or four classes, with thirty classes being the most that the authors are aware of (Train, 2008). By using a more structured support, such as a grid with equal or unequal intervals, models with several hundred classes can be estimated quite feasibly.

The limitation is largely computational. For the same number of classes, the number of parameters that needs to be estimated decreases in going from unstructured to structured supports (and within the models with structured supports, the grid with equal intervals has fewer parameters than the analogous grid with unequal intervals). While this implies greater flexibility in the shapes that can be mimicked by the first specification, it also imposes a higher computational cost. As a result, LCCMs with unstructured supports take longer to estimate than analogous LCCMs with structured supports and the same number of classes. More importantly though, as the number of classes increases, in order to obtain results that are behaviorally meaningful, the estimation process for LCCMs with unstructured supports needs significantly greater supervision. Subsequent sections will describe the process in greater detail, but it merits mentioning here that for both unstructured and structured supports, model parameters are usually estimated by maximizing the likelihood function, either directly through gradient-based optimization routines or, as is more common, using the expectation maximization (EM) algorithm (Dempster et al., 1977). The likelihood function for both LCCMs is not globally concave and has multiple local maxima. Neither gradient-based optimization routines nor the EM algorithm guarantee convergence to the global maximum. As the number of classes increases, the estimation process becomes increasingly sensitive to starting values. Models are typically built incrementally, where estimates from models with fewer classes are used to determine starting values.

In the case of LCCMs with structured supports, we argue that the estimation routine is not as sensitive to starting values, and the selection of good starting values itself is much easier. Take, for example, the two-dimensional coefficient space shown in Figure 1. Assume that the estimated mass points for a fifteen-class LCCM where the coefficient space is specified as a grid with equal intervals, with five points along the first dimension and three points along the second, are as shown in Figure 1(B). The location of each of the fifteen points can be specified as a function of four model parameters ($\alpha_1$, $\alpha_2$, $\delta_1$ and $\delta_2$), when the analogous fifteen-class LCCM with an unstructured support would require thirty model parameters. In general, estimation routines tend to be more robust when there are fewer model parameters that need to be estimated. Suppose now that the analyst wishes to estimate a model where the number of points along the second dimension is increased to four. One can assume with some certainty that increasing the number of mass points along a particular dimension will not significantly change the location of the grid boundary along that dimension. And therefore, estimates for the fifteen-class model can be used as starting values for the twenty-class model. Contrast this with the case of the six-class LCCM with an unstructured support, shown in Figure 1(A). One can similarly assume with some certainty that in going from six to seven classes, the location of the mass points for the six original classes will not change significantly, and estimates from the six-class LCCM can serve as starting values for the location of the mass points for six of the seven classes for the new model. However, estimation results from the six-class LCCM offer little insight on the location of the mass point corresponding to the seventh class, and in initializing starting values for the seventh class, different heuristics can be employed. One could use the mean of the parameter estimates for the six-class

LCCM as starting values for the seventh class. Or estimates from a simpler model, such as a multinomial logit. Train (2008) suggests partitioning the dataset into as many subsets as there are classes, estimating a multinomial logit model on each of these subsets, and using these estimates as starting values for each of the class-specific choice models. As mentioned before, estimation results for the seven-class LCCM may vary, at times significantly, based on where the estimation routine is initialized. For a given number of classes, the analyst typically estimates several models with different starting values, and selects the optimal model based on a combination of goodness of fit and behavioral interpretation. The process is repeated each time the number of classes is increased, and can prove to be quite cumbersome. Owing to these difficulties, it is often easier to estimate models with a much greater number of classes when using models with structured supports, and the resulting probability distributions are ultimately much more flexible than corresponding distributions from models with unstructured supports.

The number of classes for each of the three model specifications must be specified by the analyst prior to model estimation. Models with more classes can better approximate more flexible distributions, but they may risk overfitting the data. As mentioned previously, the appropriate number of classes is usually determined by comparing in-sample statistical measures of fit, such as the Bayesian Information Criterion (BIC) or the Akaike Information Criterion (AIC), out-of-sample measure of fit, such that the log-likelihood at convergence for a holdout sample, and the attendant behavioral interpretation. In the case of nonparametric models with a large number of parameters, in-sample measures such as the BIC and AIC, which penalize models for the number of parameters, are heavily loaded against these models. Therefore, it is almost imperative that these models be able to demonstrate greater out-of-sample predictive ability. For a more detailed recent discussion on model selection, the reader is referred to Hooten and Hobbs (2015).

Having specified the vector of taste coefficients $\boldsymbol{\beta_s}$ using one of the three ways described in preceding paragraphs, equation (5) may be combined iteratively over all decision-makers in the sample to yield the following expression for the population log-likelihood function:

$$\log L(\mathbf{y}|\mathbf{x},\mathbf{z};\mathbf{B},\boldsymbol{\gamma}) = \sum_{n=1}^{N} \log \left( \sum_{s=1}^{S} \gamma_s f_y(\mathbf{y_n}|\boldsymbol{\beta_{ns}}) \right) \qquad (9)$$

where N is the size of the sample population. Depending upon how $\boldsymbol{\beta_s}$ is specified, the appropriate equation can be substituted in the above expression, and theoretically, the unknown model parameters can be estimated by maximizing the resulting log-likelihood equation using gradient-based optimization routines, such as the BFGS method. In practice, as pointed out by Train (2008), due to the functional form (i.e. the logarithm of a sum), gradient-based optimization routines do not perform very well for nonparametric discrete mixture models such as LCCMs, due largely to three reasons: (1) the gradient does not always have a tractable analytical expression, and numerical approximation can lead to significant increases in estimation times because of the large number of model parameters that need to be estimated; (2) the likelihood function can be extremely flat, resulting in a badly-behaved Hessian that may potentially not be invertible in certain neighborhoods of the coefficient space; and (3) the likelihood function is not globally concave and the optimizer is more prone to getting stuck in neighborhoods where the function is not well approximated by a quadratic expression.

For most finite mixture models, including LCCMs, the EM algorithm proves to be a more viable alternative for model estimation (see McLachlan and Peel, 2004 for a detailed discussion). The EM algorithm is an iterative method for determining maximum likelihood estimates for models with latent variables. In the context of discrete choice models, Bhat (1997) was the first to propose the use of the EM algorithm for the estimation of LCCMs with unstructured support. Later, Train (2008) adapted Bhat's (1997) approach to estimate LCCMs with fixed support points. For a general discussion of the EM algorithm in the context of discrete choice models, the reader is also referred to Train (2009). EM algorithms are particularly attractive for finite mixture models because the optimization problem for these models can usually be reduced to the optimization of simpler constituent functions that are computationally more tractable than the original composite objective function. As the next section demonstrates, this characteristic holds true for the particular framework presented here as well. In terms of convergence rate, the EM algorithm is slower than gradient-based optimization routines, particularly in neighborhoods close to the local maximum. However, the EM algorithm demonstrates greater stability, "such that overstepping in areas of the parameter space distant from the likelihood maximum does not occur" (Ruud, 1991). Due to both the analytical tractability and the numerical stability of the optimization algorithm, the EM algorithm is the recommended method for model estimation.

## 4. An Expectation Maximization (EM) Algorithm for Model Estimation

We being the description of the EM algorithm by formulating the joint probability of observing the vector of dependent observable variables **y** and the vector of latent variables **q**, assuming as if the latent variables can be observed:

$$f_{Y,Q}(\mathbf{y}, \mathbf{q}|\mathbf{x}, \mathbf{z}; \mathbf{B}, \mathbf{\gamma}) = \prod_{n=1}^{N}\prod_{s=1}^{S}[P(q_{ns}=1|\mathbf{\gamma})]^{q_{ns}} \prod_{t=1}^{T_n}\prod_{j\in C_{nt}}[P(y_{ntj}=1|q_{ns}=1, \mathbf{x_{nt}}; \mathbf{\beta_s})]^{y_{ntj}q_{ns}} \tag{9}$$

The joint probability distribution function can be thought of as the likelihood function for the data if all of the variables could be observed and measured by the analyst. Taking the logarithm, we get the following form for the complete log-likelihood function:

$$\log L(\mathbf{y}, \mathbf{q}|\mathbf{x}, \mathbf{z}; \mathbf{B}, \mathbf{\gamma}) = \sum_{n=1}^{N}\sum_{s=1}^{S} q_{ns} \log P(q_{ns}=1|\mathbf{\gamma})$$

$$+ \sum_{s=1}^{S}\sum_{n=1}^{N}\sum_{t=1}^{T_n}\sum_{j\in C_{nt}} y_{ntj} q_{ns} \log P(y_{ntj}=1|q_{ns}=1, \mathbf{x_{nt}}; \mathbf{\beta_s}) \tag{10}$$

The EM algorithm first finds the expected value of the complete log-likelihood function with respect to the latent variables, given the observable variables and the current estimates for the unknown model parameters, denoted by the superscript $d-1$, where d is the current iteration of the algorithm. The evaluation of this expectation is called the E-step of the algorithm. In our case, the E-step can be reduced to taking the expectation of the latent variable $q_{ns}$, which may be calculated as follows:

$$E[q_{ns}|\mathbf{y_n}, \mathbf{x_n}, \mathbf{z_n}; \mathbf{B}^{(d-1)}, \mathbf{\gamma}^{(d-1)}] = P(q_{ns}=1|\mathbf{y_n}, \mathbf{x_n}, \mathbf{z_n}; \mathbf{\beta_s}^{(d-1)}, \mathbf{\gamma}^{(d-1)})$$

$$\Rightarrow q_{ns}^{(d)} = \frac{f_y(\mathbf{y_n}|q_{ns}=1, \mathbf{x_n}; \mathbf{\beta_s}^{(d-1)}) P(q_{ns}=1|\mathbf{\gamma}^{(d-1)})}{\sum_{s'=1}^{S} f_y(\mathbf{y_n}|q_{ns'}=1, \mathbf{x_n}; \mathbf{\beta_{s'}}^{(d-1)}) P(q_{ns'}=1|\mathbf{\gamma}^{(d-1)})} \tag{11}$$

where $q_{ns}^{(d)}$ is the expectation of the latent variable in the $d^{th}$ iteration of the E-step. The reader should note that equations (1) and (7) can be substituted in equation (11) to obtain estimates for the expectation. The second step of the EM algorithm, also called the M-step, maximizes the expectation of the complete log-likelihood function for each of the unknown model parameters, using outputs from the E-step. In our case, the M-step can be reduced to maximizing the following objective functions:

$$\boldsymbol{\gamma}^{(d)} = \underset{\boldsymbol{\gamma}}{\operatorname{argmax}} \sum_{n=1}^{N} \sum_{s=1}^{S} q_{ns}^{(d)} \log \gamma_s \tag{12}$$

$$\mathbf{B}^{(d)} = \underset{\mathbf{B}}{\operatorname{argmax}} \sum_{s=1}^{S} \sum_{n=1}^{N} \sum_{t=1}^{T_n} \sum_{j \in \mathbf{C}_{nt}} y_{ntj} q_{ns}^{(d)} \log P(y_{ntj} = 1 | q_{ns} = 1, \mathbf{x}_{nt}; \boldsymbol{\beta}_s) \tag{13}$$

The EM algorithm iterates between the E-step and the M-step, until some convergence criterion is satisfied. Each iteration is guaranteed to increase the likelihood function, as given by equation (8), and the algorithm is guaranteed to converge to a local maximum of the function. The convergence criterion is usually defined as a sufficiently small change in the likelihood function, though some studies have used changes in the model parameter values or the gradient as stopping conditions too (see, for example, Abbi et al., 2008).

The EM-algorithm is a deterministic algorithm that, for a given set of starting values, will always converge to the same estimates. In practice, a good set of starting values is critical. A number of heuristics have been used in the literature for selecting starting values for the model parameters (see, for example, Biernacki et al., 2003), such as using multiple shorter runs of the algorithm, but by far the most commonly employed approach is random initialization. In our case, as mentioned before, starting values will be selected based on estimation results for models with fewer classes.

Depending upon the assumptions about the class membership and class-specific choice models, equations (12) and (13) result in different objective functions. The advantage of the EM algorithm derives from the fact that the objective functions given by equations (12) and (13) are usually easier to optimize than the composite likelihood function given by equation (8). For example, updates for the class membership model parameters have a closed form solution that can be derived analytically as follows:

$$\gamma_s^{(d)} = \frac{\sum_{n=1}^{N} q_{ns}^{(d)}}{\sum_{n=1}^{N} \sum_{s=1}^{S} q_{ns}^{(d)}} \tag{14}$$

Similarly, if the class-specific choice model is specified as a multinomial logit model, then equation (13) requires the analyst to maximize weighted logit models, with the weights given by the expectation of the latent variable. For weighted logit models, the gradient can be computed analytically and the objective function is globally concave. The difference between a traditional LCCM with an unstructured support and the framework described here lies in the form of equation (13). Staying with the example where the class-specific choice model is specified as a multinomial logit model, in the case of an LCCM with an unstructured support, equation (13) can be maximized independently for each column of the matrix of taste coefficients **B**:

$$\boldsymbol{\beta}_s^{(d)} = \underset{\boldsymbol{\beta}_s}{\operatorname{argmax}} \sum_{n=1}^{N} \sum_{t=1}^{T_n} \sum_{j \in \mathbf{C}_{nt}} y_{ntj} q_{ns}^{(d)} \log \left( \frac{\exp(\boldsymbol{\beta}_s^T \mathbf{x}_{ntj})}{\sum_{j' \in \mathbf{C}_{nt}} \exp(\boldsymbol{\beta}_s^T \mathbf{x}_{ntj'})} \right) \tag{15}$$

In the case of an LCCM with a structured support, the class-specific taste coefficients are no longer independent and, depending upon whether the support has been specified as a grid with equal or unequal intervals, one of the following equations applies:

$$\boldsymbol{\alpha}^{(d)}, \boldsymbol{\delta}^{(d)} = \underset{\boldsymbol{\alpha},\boldsymbol{\delta}}{\operatorname{argmax}} \sum_{s=1}^{S} \sum_{n=1}^{N} \sum_{t=1}^{T_n} \sum_{j \in C_{nt}} y_{ntj} q_{ns}^{(d)} \log \left( \frac{\exp\left((\boldsymbol{\alpha} + \mathbf{H_s}\boldsymbol{\delta})^T \mathbf{x_{ntj}}\right)}{\sum_{j' \in C_{nt}} \exp\left((\boldsymbol{\alpha} + \mathbf{H_s}\boldsymbol{\delta})^T \mathbf{x_{ntj'}}\right)} \right) \quad (16)$$

$$\boldsymbol{\lambda}^{(d)} = \underset{\boldsymbol{\lambda}}{\operatorname{argmax}} \sum_{s=1}^{S} \sum_{n=1}^{N} \sum_{t=1}^{T_n} \sum_{j \in C_{nt}} y_{ntj} q_{ns}^{(d)} \log \left( \frac{\exp\left(\sum_{k=1}^{K} (\mathbf{h_{sk}^T} \boldsymbol{\lambda_k}) x_{ntjk}\right)}{\sum_{j' \in C_{nt}} \exp\left(\sum_{k=1}^{K} (\mathbf{h_{sk}^T} \boldsymbol{\lambda_k}) x_{ntj'k}\right)} \right) \quad (17)$$

In all three cases, the objective function is a weighted logit function that can be maximized quite efficiently using standard unconstrained non-linear optimization methods. However, for reasons mentioned previously, and as will be demonstrated in subsequent sections, models with a much greater number of classes can be estimated quite feasibly using equations (16) and (17) than would be possible using equation (15).

In cases where the analyst wishes to impose prior constraints on the parameter space, the objective function may be maximized using constrained non-linear optimization methods, such as the L-BFGS-B algorithm (Byrd et al., 1995; Zhu et al., 1997; and Morales and Nocedal, 2011). For example, when estimating travel mode choice models, taste coefficients denoting sensitivities to travel times and costs are frequently constrained to be non-positive. The imposition of constraints does not complicate the estimation procedure, because the objective function is still a weighted logit function with an easily calculable analytical expression for the gradient.

All models presented in this study were estimated in Python, and the M step of the EM algorithm was executed using an implementation of the BFGS algorithm for unconstrained optimization and the L-BFGS-B algorithm for constrained optimization, using implementations contained in the SciPy library (Jones et al., 2001). Standard errors are calculated using the methods described in Ruud (1991). To encourage other researchers to use these models, Python scripts for estimation have been provided as part of the online supplementary material.

## 5. Monte Carlo Experiments

The objective of this section is to use a series of Monte Carlo experiments to evaluate the ability of the model framework and estimation routine described in Sections 3 and 4 to recover the true taste coefficient distributions for different data generating processes.

### 5.1 Monte Carlo Experiment I: Univariate distributions

The first experiment is concerned with the analysis of univariate taste parameter distributions. For this purpose, we construct a hypothetical model of travel mode choice behavior where, for a given trip, a decision-maker can choose between four travel modes: walk, bike, car and public transit. The utility of travel mode j as perceived by decision-maker n during observation t, denoted $u_{ntj}$, is specified as a linear function of the travel time and cost incurred by that travel mode, denoted $tt_{ntj}$ and $cost_{ntj}$, respectively, and some stochastic component assumed to be i.i.d. Gumbel across decision-makers, observations and travel modes, denoted $\varepsilon_{ntj}$:

$$u_{ntj} = asc_j + \beta_{tt,n} tt_{ntj} + \beta_{cost} cost_{ntj} + \varepsilon_{ntj} \qquad (18)$$

where $asc_j$ denote the alternative-specific constant corresponding to travel mode j, $\beta_{tt,n}$ denotes the decision-maker's sensitivity to travel times, assumed to vary randomly across decision-makers in the sample population, and $\beta_{cost}$ denotes the sensitivity to cost, assumed to be the same across decision-makers in the sample population. We simulate three datasets where the population distribution for sensitivity to travel times is assumed to have a normal distribution, a lognormal distribution, and a mixture of two normal distributions, respectively. Decision-makers are assumed to be utility maximizing in that, for a given observation, they choose the travel mode that offers the greatest utility. For each of the three distributions, a dataset with 1000 pseudo-observed individuals is generated, where each individual is pseudo-observed to have made 10 choices, making for a total of 10,000 observations. For more details on the simulation process, refer to Appendix A.

**Figure 2:** A plot where the dashed lines denote the marginal cumulative distribution functions for value of time used to generate the data, and the step lines denote the analogous marginal cumulative distribution function estimated by an LCCM where the coefficient space is specified as a grid with unequal intervals

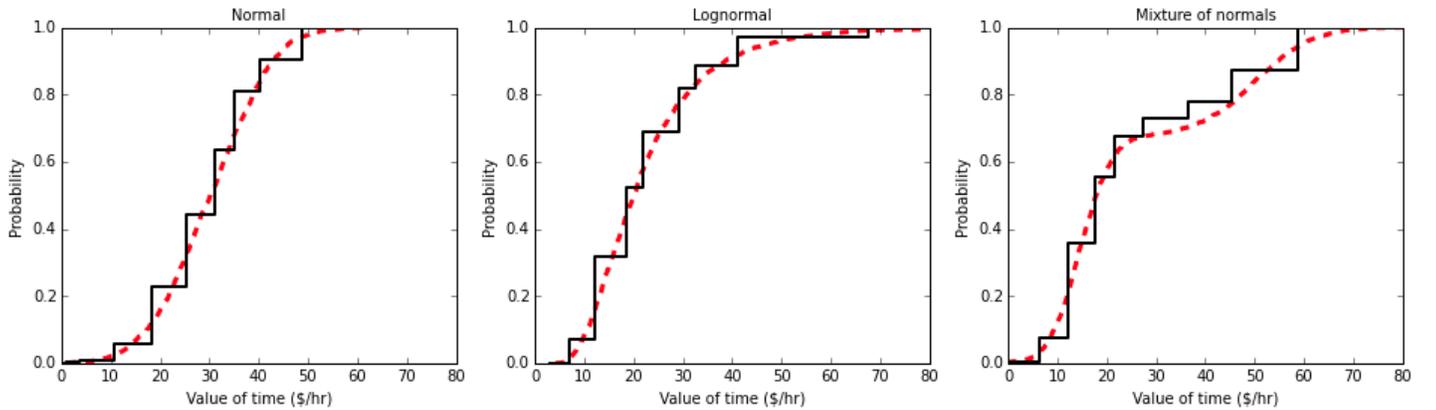

**Table 1:** True values for the fixed coefficients, and corresponding estimates from LCCMs with unequal intervals grids, for Monte Carlo experiments I and II

| Fixed coefficient | Experiment I | | | | | | Experiment II | |
|---|---|---|---|---|---|---|---|---|
| | Normals | | Lognormals | | Mixture of Normals | | | |
| | True value | LCCM estimate | True value | LCCM estimate | True value | LCCM estimate | True value | LCCM estimate |
| Constant – Bike | -3.50 | -3.54 | -3.50 | -3.51 | -3.50 | -3.59 | -3.50 | -3.41 |
| Constant – Car | 2.50 | 2.71 | 2.50 | 2.61 | 2.50 | 2.47 | -1.50 | -1.55 |
| Constant – Public transport | 0.50 | 0.57 | 0.50 | 0.56 | 0.50 | 0.48 | -2.00 | -2.01 |

The datasets are used to estimate LCCMs with varying number of mass points located at equal and unequal intervals along the travel time dimension. Figure 2 plots the distribution for value of time, as represented by the ratio of the coefficients $\beta_{tt}$ and $\beta_{cost}$, as estimated by a nine-class model with unequal intervals, and the true distributions used to generate the datasets. The relative similarity between the two plots across the three datasets attests to the ability of the model to accurately recover the true distributions. Table 1 lists estimates for the fixed coefficients. Across all specifications for the data generating process, the LCCM is able to recover the true values for the fixed coefficients as well to a high degree of accuracy. For the sake of brevity, we exclude analogous results for the model with equal intervals, and for models with different numbers of classes. In general, the degree of accuracy of the estimated distribution is higher for models with unequal intervals. This is not surprising, since the model with equal intervals is a restricted version of the model with unequal intervals. The degree of accuracy expectedly increases with the number of classes. The benefit of the approach is that it does not require the analyst to make any prior assumptions about the shape of the distribution. As illustrated by Figure 2, for each of the datasets, the same nine-class model specification is able to estimate three very different distributions.

### 5.2. Monte Carlo Experiment II: Bivariate normal distribution

In this section, we evaluate the ability of the proposed framework to recover a bivariate normal distribution. For the sake of consistency, we will be staying with the example of a hypothetical model of travel mode choice behavior constructed in the previous section. In this case, the travel time incurred by each travel mode is decomposed into in-vehicle and out-of-vehicle travel time, denoted ivtt and ovtt, respectively:

$$u_{ntj} = asc_j + \beta_{ivtt,n} ivtt_{ntj} + \beta_{ovtt,n} ovtt_{ntj} + \beta_{cost} cost_{ntj} + \varepsilon_{ntj} \qquad (19)$$

where $\beta_{ivtt,n}$ and $\beta_{ovtt,n}$ denote the decision-maker's sensitivity to in-vehicle and out-of-vehicle travel times, respectively, each of which are assumed to vary randomly across decision-makers in the sample population. The variable and coefficient distributions are listed in Appendix A. The population distribution for sensitivity to in-vehicle and out-of-vehicle travel times are assumed to be bivariate. A dataset with 1000 pseudo-observed individuals is generated, where each individual is pseudo-observed to have made 10 choices, making for a total of 10,000 observations.

The dataset is subsequently used to estimate LCCMs with varying number of mass points located at unequal intervals along the in-vehicle and out-of-vehicle travel time dimensions. Figure 3 plots the true bivariate cumulative distribution for value of in-vehicle and out-of-vehicle travel time used to generate the datasets, and the analogous distribution estimated by an eighty-one-class model, with nine mass points each along the in-vehicle and out-of-vehicle travel time dimensions. Similarly, Figure 4 plots the true and estimated marginal cumulative distributions for value of in-vehicle and out-of-vehicle travel time for the same model. The relative similarity between the true and estimated distributions attests once again to the ability of the model to accurately recover the true distributions. Note further that the LCCM, by virtue of being multivariate, is able to identify the underlying correlation between sensitivities to in-vehicle and out-of-vehicle travel time. The true underlying correlation was assumed to be 0.28, and the correlation estimated by the LCCM is 0.27.

**Figure 3:** A plot showing the true cumulative distribution function used to generate the data and the analogous cumulative distribution function estimated by an LCCM where the coefficient space was specified as a grid with unequal intervals

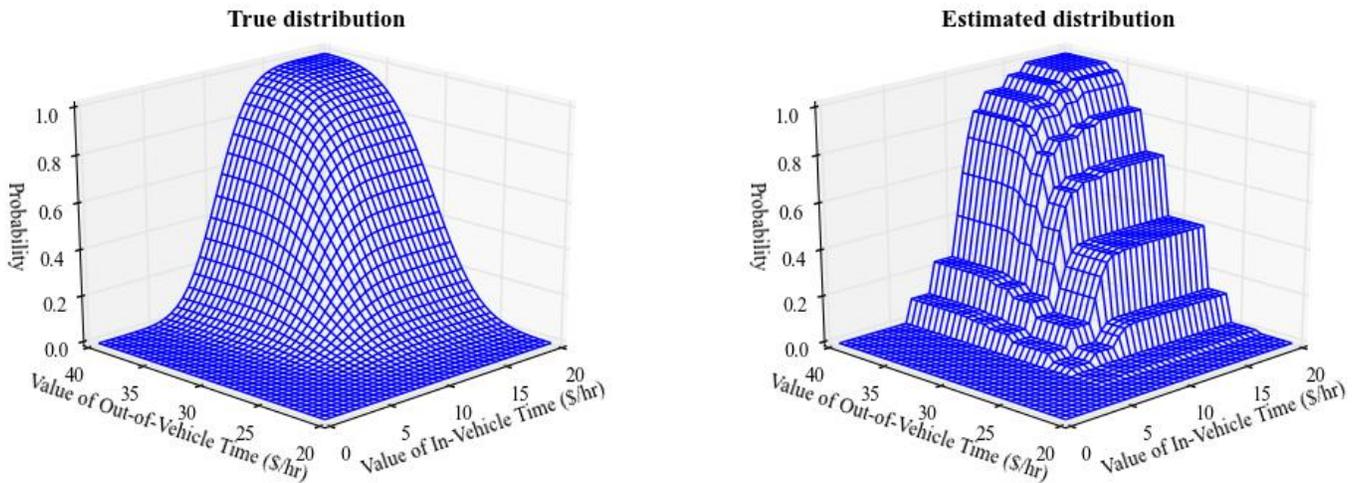

**Figure 4:** A plot where the dashed lines denote the marginal cumulative distribution functions for the two taste coefficients used to generate the data, and the step lines denote the analogous marginal cumulative distribution functions estimated by an LCCM where the coefficient space is specified as a grid with unequal intervals

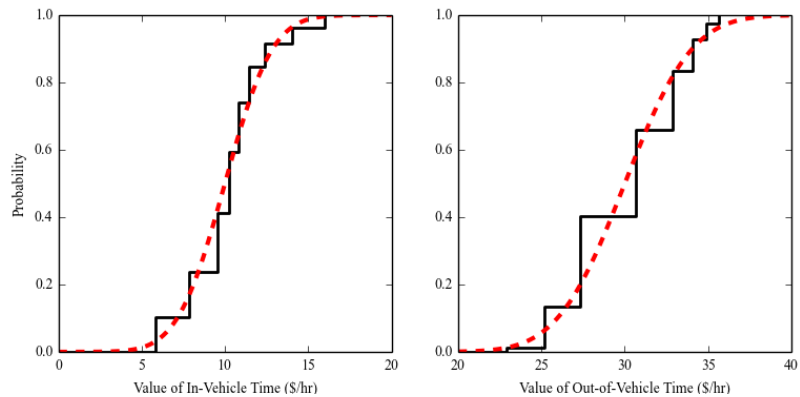

## 5.3. Monte Carlo Experiment III: Multivariate, multimodal, truncated distribution

As a final stress test, we assess the ability of the proposed model framework and estimation routine to recover taste parameter distributions from a comparatively complicated data generating process. We generate synthetic choice data for a hypothetical stated choice experiment that is concerned with the elicitation of preferences for private vehicles. Our synthetic sample consists of 1,000 individuals, who are pseudo-observed to complete ten choice tasks each, resulting in 10,000 total observations. Each choice task features three unlabeled alternatives, which are characterized by five attributes: purchase price, operating cost, a binary variable indicating if the car's powertrain is electric or not, a binary variable indicating if the car's powertrain is hybrid or not, and a binary variable indicating if the car is a premium automotive brand or not. The utility of each alternative is assumed to be linear in each of these five attributes:

$$u_{ntj} = \boldsymbol{\beta_n^T} \boldsymbol{x_{ntj}} + \varepsilon_{ntj} \tag{19}$$

where $\boldsymbol{\beta_n}$ is a $(5 \times 1)$ vector of taste coefficients and $\boldsymbol{x_{ntj}}$ is a $(5 \times 1)$ vector of covariates; and $\varepsilon_{ntj}$ is the stochastic component, assumed to be i.i.d. Gumbel across decision-makers, observations and alternatives. The taste coefficients $\boldsymbol{\beta_n}$ are drawn randomly for each pseudo-observed individual from a mixture of three multivariate truncated normal distributions. For more details on the variable and coefficient distributions, the reader is referred to Appendix A.

The data are used to estimate a 3125-class model, with 5 mass points along each dimensions and unequal intervals between successive mass points. The performance of the model is benchmarked against a hierarchical Bayesian (HB) logit model, which is known to perform well at recovering correlated taste parameter distributions (e.g. Scarpa et al., 2008). A comprehensive discussion of the HB logit model is provided in Ben-Akiva et al. (2015). Given the parametric nature of the HB logit model, we are required to select a taste parameter distribution prior to the estimation of the model. For the purpose of this simulation study, we assume that the random taste parameters are drawn from a multivariate normal distribution, whereby draws for the taste parameters pertaining to the attributes purchase price and operating cost are exponentiated and multiplied by negative one to assure that the random taste parameters are strictly negative. Consequently, the estimated taste parameter distribution is a multivariate normal-lognormal distribution. Appendix B formally develops the specification of the HB logit model and outlines the estimation approach.

**Figure 5:** A plot showing the true marginal probability mass functions for the five taste coefficients used to generate the data, and the analogous functions as estimated by an LCCM where the coefficient space is specified as a grid with unequal intervals, and an HB logit model with a multivariate normal-lognormal distribution

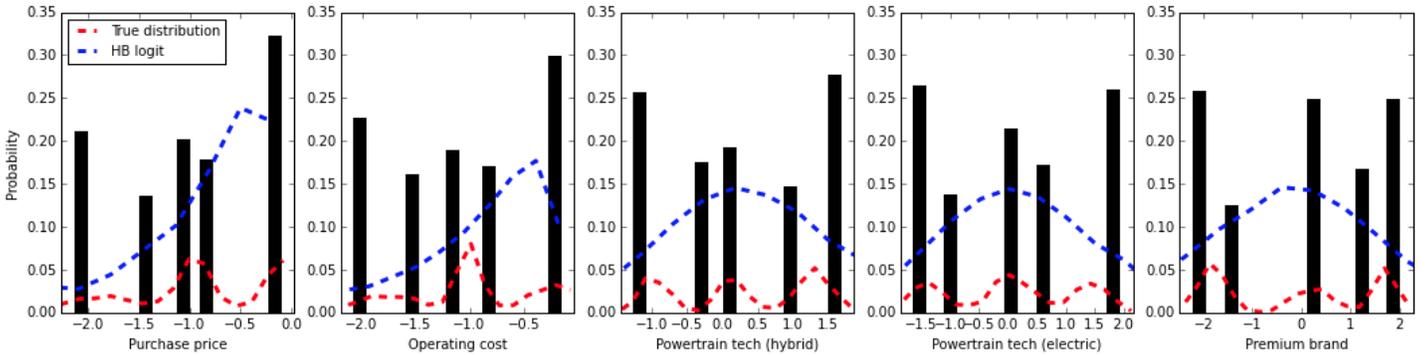

**Figure 6:** True covariance matrix, and estimates for the same from the proposed LCCM framework

| | | True values | | | | | | LCCM estimates | | | |
|---|---|---|---|---|---|---|---|---|---|---|---|
| Purchase price | 0.47 | 0.38 | -0.21 | -0.70 | 0.25 | | 0.48 | 0.39 | -0.25 | -0.75 | 0.34 |
| Operating cost | 0.38 | 0.34 | -0.20 | -0.59 | 0.28 | | 0.39 | 0.45 | -0.27 | -0.69 | 0.37 |
| Powertrain tech (hybrid) | -0.21 | -0.20 | 0.91 | 0.48 | -1.39 | | -0.25 | -0.27 | 1.10 | 0.55 | -1.32 |
| Powertrain tech (electric) | -0.70 | -0.59 | 0.48 | 1.36 | -0.62 | | -0.75 | -0.69 | 0.55 | 1.62 | -0.56 |
| Premium brand | 0.25 | 0.28 | -1.39 | -0.62 | 2.25 | | 0.34 | 0.37 | -1.32 | -0.56 | 2.41 |

**Table 2:** Mean sensitivities to different alternative attributes, as assumed by the data generating process, and as estimated by different model specifications

| Model | Attribute | | | | |
|---|---|---|---|---|---|
| | Purchase price | Operating cost | Powertrain tech (hybrid) | Powertrain tech (electric) | Premium brand |
| True distribution | -0.919 | -0.968 | 0.301 | 0.073 | -0.178 |
| LCCM with unequal intervals | -0.993 | -1.072 | 0.240 | 0.050 | 0.019 |
| HB logit with multivariate normal | -1.148 | -1.190 | 0.335 | 0.105 | -0.131 |

Figure 5 compares the marginal probability mass function estimated by the proposed framework with the true underlying distribution and the distribution recovered by the HB logit model. In general, the LCCM is able to recover the trimodal nature of the distribution across each of the five dimensions, and the location of the mass points corresponds well with the location of local maxima of the true distribution. In contrast, and perhaps unsurprisingly, estimates from the HB logit do not look anything like the true distribution. Interestingly, differences in the shape of the true distribution, and the distribution estimated by the HB logit model do not lead to corresponding differences in mean sensitivities, as reported in Table 2. Both the HB logit and the LCCM are able to recover mean sensitivities to a reasonable degree of accuracy. In fact, in some cases, estimates from the HB logit are closer to the true values.

Figure 6 compares the true covariance matrix underlying the data generating process with estimates for the same from the LCCM. In general, the model is able to recover the true values to a high degree of accuracy. The signs are the same in all cases, and the root mean square error for the estimates is 0.09. The degree of accuracy indicates that a high number of mass points is not needed along each random coefficient for the model to be able to recover complex covariance structures. For the sake of brevity, we do not include analogous estimates for the HB logit model. However, it bears mentioning that the root mean square error for covariance matrix estimates from the HB logit model is 0.85. Given the model's inability to recover marginal distributions for each of the coefficients, it shouldn't be surprising that estimates for the covariance matrix are different from the true underlying values as well.

## 6. Case Study: Travel Mode Choice Behavior in the San Francisco Bay Area, United States

The objective of this section is to illustrate the benefits of the proposed framework through an application to the case of travel mode choice behavior. Data for our analysis come from the Bay Area Travel Survey (BATS) 2000, a large-scale regional household travel survey conducted in the nine county San Francisco Bay Area of California. The San Francisco Metropolitan Transportation Commission (SFMTC) has periodically sponsored BATS to provide data to support travel modeling and analysis of regional travel behavior. The target data collection period for BATS 2000 was of course the 2000 calendar year. The survey consisted of an activity-based travel diary that requested information on all in-home and out-of-home activities over a two-day period, including weekday and weekend pursuits. In all, more than 15,000 households participated in the survey. More information on the raw data can be found in Morpace International, Inc. (2002).

Individual activities and trips are processed into home-based tours. Depending on the primary destination, home-based tours can subsequently be classified as mandatory or non-mandatory tours. A mandatory tour includes a trip to a place of employment or education; a non-mandatory tour does not. Travel demand analysts typically estimate separate mode choice models for each tour type. For the purpose of our analysis, we have limited attention to mandatory tours. In all, travel diary data for 30,166 tours made by 17,700 individuals are used for model estimation, and travel diary data for an additional 3,440 tours made by 1,967 individuals are used for model validation. For each tour, four feasible travel mode alternatives are defined: car, public transit, bike, and walk. We used travel skims generated by the San Francisco County Transportation Authority's travel demand model to construct for each tour the feasible choice set and the level-of-service attributes of each of the travel modes contained within the choice set.

The data are used to estimate different multinomial logit, mixed logit and LCCM specifications. Our baseline for comparison is a multinomial logit model with the following utility specification:

$$u_{ntj} = asc_{nj} + \sum_{k=1}^{4} \beta_{time_k} time_{k,ntj} + \beta_{cost,n} cost_{ntj} + \varepsilon_{ntj} \tag{20}$$

$$asc_{nj} = \beta_{const,j} + \sum_{k=1}^{5} \beta_{age_k,j} age_{nk} + \beta_{male,j} male_n + \beta_{carless,j} carless_n \tag{21}$$

$$\beta_{cost,n} = \beta_{cost\_mean} + \beta_{cost\_inc} inc_n \tag{22}$$

where $asc_{nj}$ is the alternative-specific constant, specified itself as a linear function of age, gender, employment and car ownership, such that $age_{n1}$ is a binary variable that equals one if the decision-maker's age is 12 years or less, $age_{n2}$ equals one if age is 13-16 years, $age_{n3}$ equals one if age is 25-44 years, $age_{n4}$ equals one if age is 45-64 years, and $age_{n5}$ equals one if age is 65 years or older; $male_n$ is a binary variable that equals one if the decision-maker's gender is male; and $carless_n$ is a binary variable that equals one if the number of cars owned by the household corresponding to decision-maker n is less than the number of workers in that household. Gender was found to have a significant effect on the utility of bicycling alone, and the baseline logit specification includes gender only in utility specification for bicycling. Similarly, $time_{k,ntj}$ denotes the $k^{th}$ travel time component for alternative j, where there are four travel time components for each mode: in-vehicle time, waiting time, walking time and bicycling time. Sensitivity to travel cost, $cost_{ntj}$, is specified itself as a function of the decision-maker's household income, $inc_n$. And finally, for the baseline model, the stochastic component $\varepsilon_{ntj}$ is assumed to be i.i.d.

Gumbel with location zero and scale one across alternatives, observations and decision-makers. The specification is loosely based on the work tour travel mode choice model component of the San Francisco County Travel Demand Forecasting Model developed by the San Francisco County Transportation Authority (Cambridge Systematics, 2002).

The LCCM specifications, with equal and unequal intervals, allow the three alternative-specific constants $\beta_{const,j}$ and the four travel time coefficients $\beta_{time_k}$ to vary across classes (one of the alternative-specific constants is constrained to be zero, to enable identification). By holding the travel cost coefficients constant across classes, the model is implicitly estimated in willingness to pay space. By allowing the alternative-specific constants to vary across classes, we employ a discrete approximation to continuous mixture models with heteroskedastic error components. Similarly, by allowing the travel time coefficients to vary across classes, we employ a discrete approximation to continuous mixture models with random taste heterogeneity. By specifying utility as an additional function of demographic variables, we allow systematic taste heterogeneity as well.

We contrast the performance of the proposed LCCM with the analogous continuous mixture models with heteroskedastic error components, systematic taste heterogeneity and randomly distributed taste parameters, such that:

$$\varepsilon_{ntj} = \eta_{nj} + \nu_{ntj}, \eta_{nj} \sim \text{Normal}(0, \sigma_j^2) \text{ and } \nu_{ntj} \sim \text{Gumbel}(0,1) \quad (23)$$

$$\beta_{time_k} \sim \text{Normal}(\psi_k, \phi_j^2) \text{ or } -\beta_{time_k} \sim \text{Lognormal}(\psi_k, \phi_j^2) \quad (24)$$

where $\sigma_j$, $\psi_k$ and $\phi_j$ are additional model parameters to be estimated over the baseline specification given by equations (20)–(22). We try both univariate normal and univariate lognormal distributions for each of the travel time coefficients. We also attempt to estimate a HB logit model with a multivariate normal random taste parameter distribution in order to recover correlations between random coefficients. However, given the comparatively large sample size, we are unable to estimate such a model using the inference approach outlined in Appendix B. While Bayesian methods have been argued to perform better than maximum simulated likelihood methods at recovering the parameters of a multivariate normal taste parameter distribution (e.g. Scarpa et al., 2008), Bayesian methods are also known to not scale well to large datasets (see Braun and McAuliffe, 2010, for a discussion in the context of discrete choice methods). It is our view that the testing and development of discrete choice methods, which allow for the identification of covariance structures between random taste parameters in large-scale applications, represents a promising avenue for future research (also see Cherchi and Guevara, 2012).

Table 3 reports summary statistics for the different multinomial logit, continuous mixtures of logit and LCCM specifications. To facilitate comparison, we enumerate for each model the number of parameters that needed to be estimated, estimation times under the convergence criterion that the difference in log-likelihood for the estimation sample between successive iterations of the EM algorithm is less than 0.1 (representing less than a 0.001% change in the log-likelihood function at convergence), the log-likelihood at convergence for the estimation and validation samples, and the BIC and AIC for the estimation samples. We tried a number of different LCCM specifications with different numbers of mass points along each of the random taste coefficients. Based on a comparison across different measures of fit, the appropriate number of mass points

**Table 3:** Summary statistics for different model specifications

| Model | Model Parameters | Estimation Time | Log-likelihood | | Bayesian Information Criterion | Akaike Information Criterion |
|---|---|---|---|---|---|---|
| | | | Estimation | Holdout | | |
| *Multinomial logit* | | | | | | |
|   Without systematic heterogeneity | 7 | <1 min | -10,521 | -1,114 | 21,110 | 21,055 |
|   With systematic heterogeneity | 28 | <1 min | -10,116 | -1,066 | 20,507 | 20,289 |
| *Continuous mixtures of logit** | | | | | | |
|   Univariate normal | 36 | 36 hours | -7,477 | -803 | 15,306 | 15,026 |
|   Univariate lognormal | 36 | 71 hours | -7,583 | -824 | 15,519 | 15,239 |
| *2048-Class LCCM (2 mass points per ASC, 4 per time coefficient)* | | | | | | |
|   Equal intervals | 2082 | 19 hours | -7,339 | -784 | 35,042 | 18,842 |
|   Unequal intervals | 2090 | 32 hours | -7,294 | -780 | 35,031 | 18,768 |

* All continuous mixture models were estimated in Python Biogeme (Bierlaire, 2016) using maximum simulated likelihood estimation with 2000 pseudo-random draws per individual. The draws were generated using the modified latin hypercube sampling (MLHS) method (Hess et al., 2006). All models (with continuous or finite mixtures) were estimated on a quad-core CPU with 3.10 GHz processors.

along the 3 alternative-specific constants was determined to be 2, and the appropriate number of mass points along the 4 travel time coefficients was determined to be 4[1]. For the sake of brevity, we enumerate summary statistics for only the final model specification.

For the full list of parameter estimates for the multinomial logit, mixed logit and LCCM specifications listed in Table 3, the reader is referred to Appendix C. Figures 7 and 8 plot the distributions for willingness to pay measures estimated by each of the models, and Table 4 reports mean and median values for the same. Table 5 reports estimates of the location of the mass points for each of the random taste coefficients for the two LCCM specifications. Over following paragraphs, we review some of the major findings from our analysis.

With regards to goodness-of-fit, measures such as the BIC and the AIC that penalize models for the number of parameters tend to favor more parsimonious specifications. If we were to limit our attention to the BIC and the AIC, the mixed logit model with univariate normal distributions would be the preferred model specification. This is not surprising. The nonparametric LCCMs each have hundreds of model parameters, and a comparison based on measures such as BIC and AIC is strongly loaded against such model frameworks. However, a comparison based on the log-likelihood for the holdout sample leads us to a different conclusion: both LCCM specifications outperform the univariate continuous mixture specifications, indicating that these models have greater out-of-sample predictive ability. Within the LCCM specifications, the model with unequal intervals demonstrates greater out-of-sample predictive ability. Even in terms of the AIC and the BIC, the LCCM with unequal intervals performs better than the analogous model with equal intervals.

---

[1] Note that the number of parameters for this specification is equal to the sum of the number of fixed coefficients, the number of parameters needed to specify the location of the mass points corresponding to the random coefficients, and the number of classes minus one. Therefore, for the 2048-class LCCM with unequal intervals, we have 19 parameters corresponding to the fixed coefficients, 6 parameters to specify 2 mass points each along the 3 alternative-specific constants, 16 parameters to specify 4 mass points each along the 4 travel time components, and 2047 parameters to specify the probability mass for the 2048 classes, resulting in a total of 2090 parameters.

**Table 4:** Mean and median values of different travel time components for median household incomes, as estimated by different model specifications

| Model | In-vehicle time ($/hr) | | Walking time ($/hr) | | Bicycling time ($/hr) | | Waiting time ($/hr) | |
|---|---|---|---|---|---|---|---|---|
| | **Mean** | **Median** | **Mean** | **Median** | **Mean** | **Median** | **Mean** | **Median** |
| *Multinomial logit* | | | | | | | | |
| Without systematic heterogeneity | 9.0 | - | 5.3 | - | 10.4 | - | 5.7 | - |
| With systematic heterogeneity | 9.0 | - | 4.8 | - | 10.8 | - | 4.7 | - |
| *Continuous mixtures of logit* | | | | | | | | |
| Univariate normal | 15.0 | 15.0 | 25.1 | 25.1 | 60.2 | 60.2 | 11.9 | 11.9 |
| Univariate lognormal | 5.7 | 3.7 | 11.3 | 8.4 | 100.6 | 51.9 | 3.6 | 1.2 |
| *2048-Class LCCM (2 mass points per ASC, 4 per time coefficient)* | | | | | | | | |
| Equal intervals | 16.2 | 21.1 | 26.0 | 27.8 | 52.5 | 62.5 | 6.1 | 7.8 |
| Unequal intervals | 18.4 | 24.3 | 28.5 | 29.6 | 77.2 | 66.2 | 9.1 | 13.0 |

Differences in fit between different pairs of models can be explained by comparing the shapes of the estimated random taste coefficients. Figure 7 plots the value of time distributions, as estimated by LCCMs with equal intervals with 2 and 4 mass points alone each travel time coefficient (and 2 mass points along each ASC, in both cases). Note that each of the four coefficients was constrained to be negative. This is another benefit of the model framework, where these constraints can be imposed quite straightforwardly, as opposed to models with parametric distributions where the analyst must find an appropriate functional form. For the 128-class model, the location of the mass points along each of the dimensions seems to follow a pattern: individuals are either insensitive to changes in the variable (as indicated by a value of time close to zero), or are hypersensitive (as indicated by a high value of time). The 2048-class model with equal interval grids lends a more nuanced picture, with a greater fraction of the sample population exhibiting moderate sensitivities to one or more level-of-service attributes (see the corresponding marginal shares reported in Table 5).

One of the limitations of the model with equal interval grids is that, by forcing grid intervals along any one dimension to be the same, the model puts strain on the grid boundaries and limits the geometries that are permitted. The LCCM with unequal intervals relaxes this limitation. Compare, for example, the marginal distribution for the value of bicycling time, as identified by the 2048-class model with equal and unequal interval grids. The model with unequal intervals identifies one mass point corresponding to individuals with a low value of bicycling time, two mass points corresponding to individuals with moderate values of bicycling time, and one mass point corresponding to individuals with a high value of bicycling time. The analogous model with equal intervals is unable to identify the segment with a high value of bicycling time.

The wide variety of distributional shapes that the framework (with equal and unequal interval grids) is able to recover demonstrates its ability to overcome some of the limitations imposed by parametric formulations. For example, in the case of the LCCM with unequal interval grids, one could perhaps argue that the value of in-vehicle time has a distribution that could be approximated by a uniform function, but the other values of time have distributions that appear harder to approximate using parametric functions. Neither the normal nor the lognormal distributions that are most frequently employed in the literature appear to be good approximations, though the normal seems more appropriate in most cases (as evidenced also by the better in-sample and out-of-sample predictive ability). Moreover, the framework did not require us to make any prior assumptions about the shape of the distributions (though it did require us to specify the number of mass points along each randomly distributed taste coefficient).

**Figure 7:** The marginal probability mass functions for the value of different travel time components for individuals with median household incomes, as estimated by LCCMs with equal intervals with 2 and 4 mass points along each travel time component (and 2 mass points along each alternative-specific constant, in both cases)

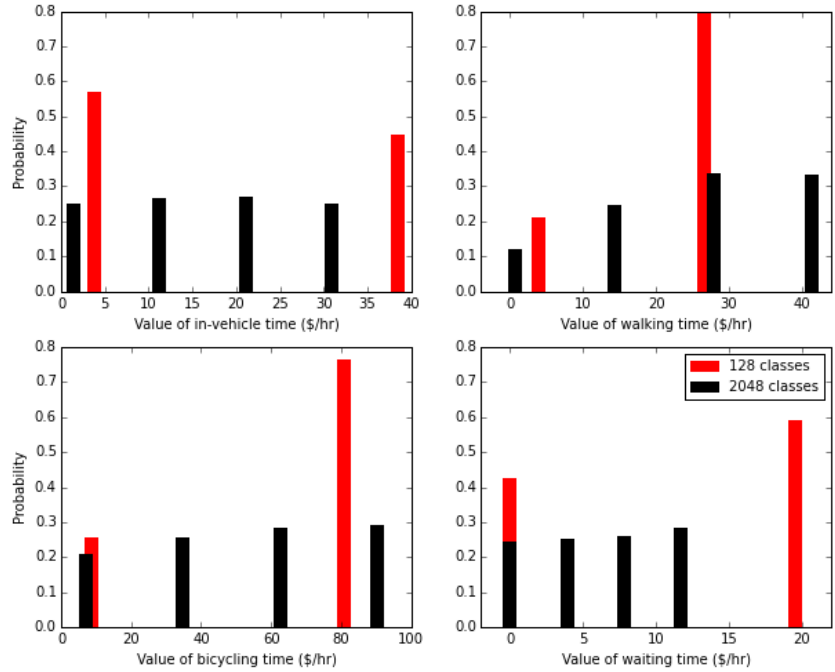

**Figure 8:** The marginal probability mass functions for the value of different travel time components for individuals with median household incomes, as estimated by 2048-class LCCMs with unequal intervals with 4 mass points along each travel time component, where the location of the mass points is constrained to be strictly non-positive for one of the cases. For comparison, the figure plots the analogous distributions estimated by the two continuous mixture models.

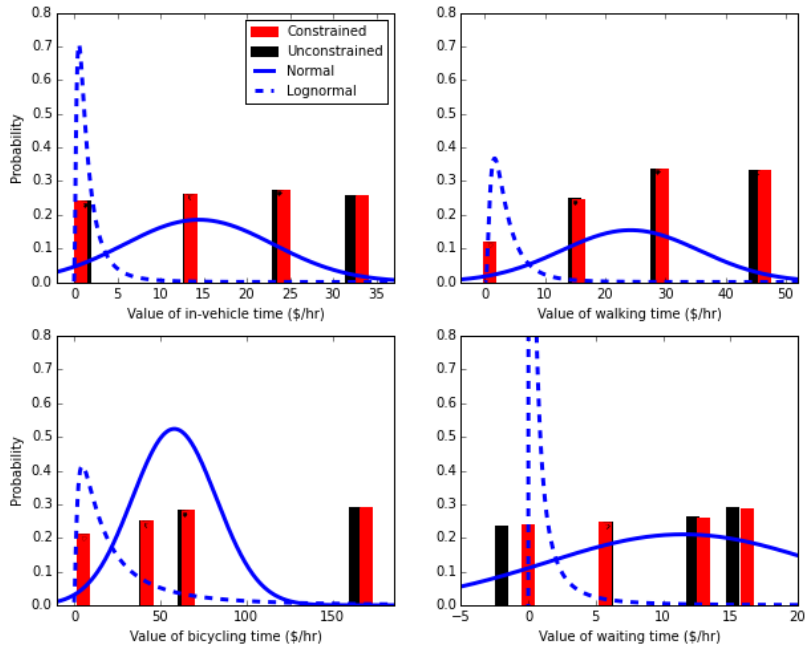

**Table 5:** Estimated location and marginal probabilities of each of the mass points along the random taste coefficients, as estimated by different LCCM specifications

| Model | 2048-class LCCM with equal intervals [a] | | | | 2048-class LCCM with unequal intervals | | | |
|---|---|---|---|---|---|---|---|---|
| | est. | t-stat | vot [b] | share | est. | t-stat | vot [b] | share |
| *Public transport specific constant* | | | | | | | | |
| Mass point 1 | -1.890 | -5.54 | - | 46% | -2.230 | -5.14 | - | 45% |
| Mass point 2 | 6.187 | 17.17 | - | 54% | 7.724 | 19.84 | - | 55% |
| *Walk specific constant* | | | | | | | | |
| Mass point 1 | -7.034 | -15.33 | - | 53% | -8.050 | -15.04 | - | 53% |
| Mass point 2 | 6.977 | 15.94 | - | 47% | 8.260 | 18.09 | - | 47% |
| *Bike specific constant* | | | | | | | | |
| Mass point 1 | -18.393 | -17.49 | - | 62% | -22.993 | -19.58 | - | 64% |
| Mass point 2 | -2.231 | -5.21 | - | 38% | -1.927 | -4.33 | - | 36% |
| *In-vehicle time (hr)* | | | | | | | | |
| Mass point 1 | -9.218 | -101.04 | 31.0 $/hr | 24% | -10.866 | -56.79 | 33.3 $/hr | 25% |
| Mass point 2 | -6.284 | - | 21.1 $/hr | 26% | -7.904 | -35.36 | 24.3 $/hr | 27% |
| Mass point 3 | -3.351 | - | 11.3 $/hr | 26% | -4.420 | -39.70 | 13.6 $/hr | 25% |
| Mass point 4 | -0.417 | -3.77 | 1.4 $/hr | 24% | -0.253 | -4.10 | 0.8 $/hr | 23% |
| *Walking time (hr)* | | | | | | | | |
| Mass point 1 | -12.330 | -35.64 | 41.4 $/hr | 32% | -15.126 | -32.96 | 46.4 $/hr | 32% |
| Mass point 2 | -8.287 | - | 27.8 $/hr | 33% | -9.643 | -35.02 | 29.6 $/hr | 33% |
| Mass point 3 | -4.244 | - | 14.3 $/hr | 24% | -5.045 | -32.70 | 15.5 $/hr | 24% |
| Mass point 4 | -0.200 | -0.56 | 0.7 $/hr | 11% | -0.234 | -1.97 | 0.7 $/hr | 11% |
| *Bicycling time (hr)* | | | | | | | | |
| Mass point 1 | -26.869 | -13.22 | 90.3 $/hr | 28% | -55.094 | -4.20 | 169.0 $/hr | 28% |
| Mass point 2 | -18.612 | - | 62.5 $/hr | 27% | -21.572 | -12.63 | 66.2 $/hr | 27% |
| Mass point 3 | -10.355 | - | 34.8 $/hr | 24% | -13.780 | -13.67 | 42.3 $/hr | 24% |
| Mass point 4 | -2.098 | -1.05 | 7.0 $/hr | 20% | -1.708 | -5.09 | 5.2 $/hr | 20% |
| *Waiting time (hr)* | | | | | | | | |
| Mass point 1 | -3.480 | -6.27 | 11.7 $/hr | 27% | -5.302 | -2.85 | 16.3 $/hr | 28% |
| Mass point 2 | -2.320 | - | 7.8 $/hr | 25% | -4.222 | -2.05 | 13.0 $/hr | 25% |
| Mass point 3 | -1.160 | - | 3.9 $/hr | 24% | -1.850 | -2.10 | 5.7 $/hr | 24% |
| Mass point 4 | 0.000 | 0.00 | 0.0 $/hr | 23% | 0.000 | 0.00 | 0.0 $/hr | 23% |

[a] For LCCMs with equal interval grids, t-stats are only reported for the location of the bounds on the grid; the location of interior points is not directly estimated, but imputed from the bounds

[b] Values of time are reported for individuals with median household incomes of $87,500 per year

Differences in the shapes of the estimated marginal distributions lead to corresponding differences in behavioral outputs, such as willingness to pay measures. First, we compare estimates from the LCCM with unequal intervals with those from the analogous LCCM with equal intervals and the continuous mixture model with univariate normals. For value of in-vehicle time, walking time and bicycling time, where the distribution estimated by the LCCM with unequal intervals is relatively symmetric, corresponding estimates from both the LCCM with equal intervals and the continuous mixture model are not very different. However, there are significant differences in the case of bicycling time. The distribution estimated by the LCCM with unequal intervals appears bimodal, with a smaller mode on the right, with roughly 28% of the sample population having an exceptionally high value of bicycling time of 170 $/hr. The LCCM with equal intervals and the continuous mixture model with univariate normals are unable to capture this bimodal shape. Both distributions are able to approximate the larger left mode, but unable to approximate the smaller right mode. And consequently, mean and median estimates from both models are smaller in comparison to the LCCM with unequal intervals.

The opposite is true for the continuous mixture model with univariate lognormals. There are significant differences in estimates for all values of time between the model and the LCCM with unequal intervals. For value of in-vehicle time, walking time and bicycling time, where marginal distributions estimated by the LCCM with unequal intervals are relatively symmetric, relative to the LCCM, the continuous mixture model significantly under predicts mean and median values of time. However, in the case of bicycling time, where the marginal distribution estimated by the LCCM with unequal intervals is bimodal, the continuous mixture model over predicts mean value of time but still under predicts the median value of time.

There are differences in estimates for other fixed taste coefficients as well, but for the sake of brevity, we do not undertake a detailed discussion here. The interested reader is pointed to Appendix C for a complete tabulation of estimation results for each of the different model specifications. Table 5 enumerates estimates for the location of mass points along each of the random taste coefficients, for the two LCCM specifications. The table demonstrates the proposed LCCM framework's ability to endogenously uncover patterns of attribute non-attendance, and to distinguish it from low sensitivities to the same attributes. For example, the 2048-class LCCM with unequal interval grids finds attribute non-attendance to be at play only with respect to waiting times, where roughly 23% of the sample population is found to be insensitive to the attribute. In the case of in-vehicle time, walking time and bicycling time, all individuals are found to consider these attributes, though a small proportion in each case has a low (but statistically significant) positive value of time.

Our findings are contrary to findings reported in the literature using more restrictive distributional frameworks, such as the confirmatory LCCM structure used by Scarpa et al. (2009) and Campbell et al. (2011) that imposes prior constraints on attribute attendance across different classes. These studies report much higher rates of non-attendance. For example, Scarpa et al. (2009) find that 90% of their sampled population may not be sensitive to costs in the context of landscape preferences. However, our findings are consistent with findings reported in the literature using more flexible distributional frameworks, such as the unrestricted LCCM with unstructured support used by Hess et al. (2013), that allow for both patterns of non-attendance and low sensitivities with respect to the same attributes. These studies report much lower rates of attribute non-attendance, and find that in many cases, higher rates of non-attendance reported by previous studies are artifacts arising from the use of more restrictive distributional frameworks.

In our case, even for waiting times, one could argue that non-attendance arises due to constraints on the coefficient space (as mentioned previously, all time coefficients are constrained to be non-positive during model estimation). Figure 8 plots the marginal distributions for different values of time as estimated by an analogous 2048-class LCCM with unequal interval grids with no sign constraints on the coefficient space. The distributions for in-vehicle time, walking time and bicycling time are expectedly identical to the corresponding distributions from the constrained model, because the constraint did not prove to be binding in each of these cases. However, for waiting time, the unconstrained model identifies a mass point in the positive orthant: per the model, 23% of the sample population has a negative value of waiting time of -2.0$/hr. The location of the mass point is statistically significantly different from zero, so the result cannot be dismissed as 'noise'. If the analyst believes that a negative value of waiting time is plausible from a behavioral perspective (as has been argued by, for example, Ory and Mokhtarian, 2005 and Cirillo and Axhausen, 2006), then empirical results from the LCCM with no sign constraints can be viewed as support for that belief. If the analyst believes that a negative value of waiting time is implausible, then estimation results from the LCCM with sign constraints can be taken as proof of attribute non-attendance.

**Figure 9:** Joint probability mass functions for different pairs of value of time for individuals with median household incomes, as estimated by the 2048-class LCCM with unequal intervals

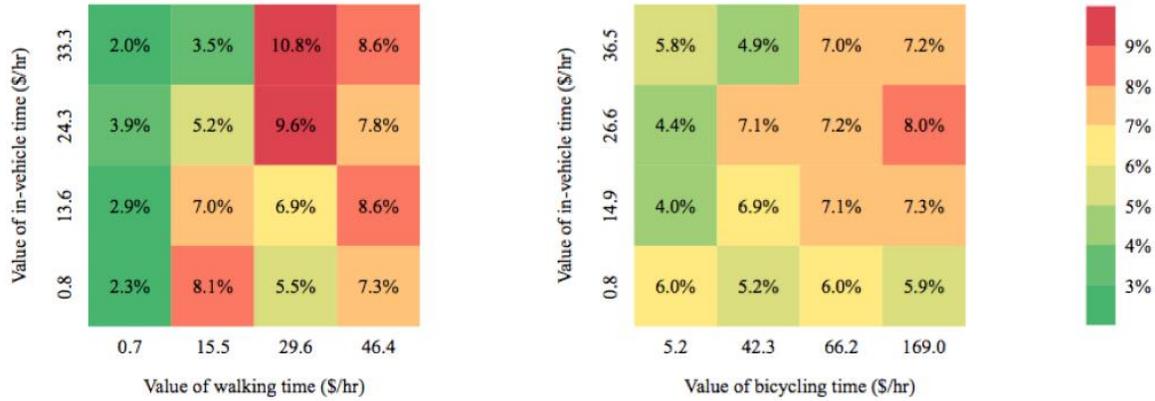

The proposed LCCM framework is able also to endogenously uncover patterns of choice set formation. Looking again at the estimation results in Table 5 for the 2048-class LCCM with unequal interval grids, 64% of the sample population has a large negative alternative-specific constant for bicycling. Individuals belonging to these classes have low predicted bicycling mode shares of approximately 0.1%, and the demand for bicycling for these classes is inelastic with respect to all level-of-service attributes, suggesting that these individuals likely do not consider bicycling as a viable option when deciding how to travel. These patterns are not as salient for other travel modes, suggesting that choice set formation may not be as much at play for these modes.

Both LCCMs, by virtue of estimating multivariate distributions, allow us to examine the joint distribution across the full coefficient space, or conditional and marginal distributions across one or more dimensions, allowing for richer analysis than would be possible using univariate parametric distributions. For example, Figure 9 plots the joint probability mass functions for different value of time pairs, as estimated by the 2048-class LCCM with unequal intervals. Individuals with low values of walking time are evenly distributed across all values of in-vehicle time, but individuals with low values of bicycling time are much more likely to exhibit either low or high values of in-vehicle time, and less likely to exhibit moderate values. In general though, we do not observe strong patterns of correlation. For example, the estimated correlation between value of in-vehicle time and value of walking time is 0.08, and the estimated correlation between value of in-vehicle time and value of bicycling time is 0.03.

And finally, estimation times for the proposed LCCM framework varied approximately from 2 hours for a 128-class model (with 2 mass points along each ASC and travel time coefficient) to as much as 5 days for a 5000-class model (with 2 mass points along each ASC and 5 mass points along each travel time coefficient). These are comparable to estimation times for continuous mixture models. However, a drawback to the framework (with equal or unequal interval grids) is the exponential increase in the number of classes for every additional random taste parameter. For example, a model with 10 random taste parameters would require a minimum of 1,024 classes and 1,033 model parameters. While the framework is appropriate for the estimation of multivariate distributions over a low-dimensional parameter space, advances in estimation methods may need to be leveraged for the estimation of these same models in higher dimensional parameter spaces, particularly with large datasets, such as parallelization through multithreading and the use of Graphics Processing Units.

# 7. Robustness of Estimation Algorithm

In this section, we evaluate the robustness of the EM algorithm for different specifications of the proposed LCCM framework. For a given model specification, we run the algorithm ten times with different starting values, and assess the stability of willingness to pay estimates across different runs. We use a subset of the BATS 2000 data used in Section 6 for our sensitivity analysis, comprising 1,031 tours made by 590 randomly sampled individuals, or roughly 3% of the full dataset. The size of the resulting dataset is comparable to most discrete choice applications in the literature, and findings from our analysis using this subset should serve as a credible indicator of the stability of model outputs that can be expected for other empirical contexts.

Table 6 reports sample means, sample standard errors and coefficients of variation for estimates of different mean values of time for households with median incomes across the ten estimation runs for each model specification. The general form of the utility function is held the same across all model specifications, and is identical to the final specification reported in Section 6. The number of mass points along each random taste coefficient and the shape of the resulting high-dimensional grid are varied across specifications. For all models, starting values for the probability mass for each class are randomly sampled from the corresponding simplex, and starting values for the fixed taste coefficients are set to the corresponding estimates from the multinomial logit model. Starting values for mass point locations for unsigned random coefficients are randomly sampled from the interval $(-10^d, 10^d)$, where $10^d$ is the smallest power of ten that is greater than the absolute value of the corresponding estimate for the parameter from the multinomial logit model. Starting values for mass point locations for signed random coefficients are randomly sampled from analogous intervals truncated appropriately at zero. In all cases, this interval is a superset of the 95% confidence interval for the same coefficient, as estimated by the multinomial logit model, and an order of magnitude larger. For example, in the case of in-vehicle time, starting values for mass point locations are sampled such that the corresponding value of in-vehicle time may vary between 0$/hr and 120$/hr, even though the 95% confidence interval for value of in-vehicle time, as estimated by the multinomial logit model, is bounded by 9.7$/hr and 14.6$/hr.

We wished also to compare the performance of the EM algorithm against more traditional gradient-based optimization methods, such as the BFGS algorithm. However, we were unable to obtain convergence for any of the model specifications reported in Table 6 using implementations of the BFGS algorithm contained in the SciPy library (Jones et al., 2001) within Python.

In general, estimates for different values of time appear to be quite stable across different model specifications, despite sampling starting values from a very broad interval. Sample standard errors are relatively small in all cases, and the coefficient of variation is considerably less than one. As one would expect, estimates from LCCMs with unequal intervals appear to be more sensitive to starting values than LCCMs with equal intervals. LCCMs with unequal intervals are unrestricted versions of analogous LCCMs with equal intervals. As model complexity increases, the likelihood function is more likely to be irregular.

**Table 6:** Sample mean, sample standard error and coefficient of variation of mean values of different travel time components for median household incomes, as estimated by different model specifications across 10 runs of the estimation algorithm with different starting values

| Model type | Model specification [a] | In-vehicle time ($/hr) | | | Walking time ($/hr) | | | Bicycling time ($/hr) | | | Waiting time ($/hr) | | |
|---|---|---|---|---|---|---|---|---|---|---|---|---|---|
| | | **Mean** | **SE** | **CV** | **Mean** | **SE** | **CV** | **Mean** | **SE** | **CV** | **Mean** | **SE** | **CV** |
| LCCM with equal interval grids | 128 classes | 11.9 | 1.3 | 0.11 | 23.3 | 4.3 | 0.18 | 29.1 | 3.0 | 0.10 | 13.2 | 1.6 | 0.12 |
| | 648 classes | 11.5 | 0.6 | 0.05 | 23.5 | 6.4 | 0.27 | 25.9 | 2.2 | 0.09 | 8.7 | 0.9 | 0.10 |
| | 2048 classes | 11.1 | 0.6 | 0.05 | 25.7 | 1.7 | 0.06 | 23.7 | 1.1 | 0.05 | 8.1 | 0.7 | 0.08 |
| | 5000 classes | 11.2 | 0.5 | 0.05 | 27.1 | 2.9 | 0.11 | 23.8 | 1.5 | 0.06 | 7.6 | 0.7 | 0.09 |
| LCCM with unequal interval grids | 128 classes | 12.2 | 1.7 | 0.14 | 24.1 | 5.0 | 0.21 | 39.9 | 13.3 | 0.33 | 16.6 | 2.1 | 0.13 |
| | 648 classes | 11.2 | 1.3 | 0.12 | 23.3 | 3.9 | 0.17 | 39.0 | 5.7 | 0.15 | 21.2 | 2.3 | 0.11 |
| | 2048 classes | 12.3 | 0.9 | 0.07 | 25.1 | 4.5 | 0.18 | 39.1 | 11.0 | 0.28 | 20.6 | 2.4 | 0.12 |
| | 5000 classes | 11.6 | 0.6 | 0.05 | 24.0 | 4.8 | 0.20 | 30.7 | 6.3 | 0.20 | 21.6 | 2.1 | 0.10 |

[a] All LCCMs have 2 mass points along each of the three alternative-specific constants, the number of mass points along each of the four travel time coefficients is varied between 2 and 5

# 8. Conclusions

The literature on discrete choice models is replete with ways to incorporate random taste heterogeneity within existing representations of disaggregate decision-making. By and large, the mixed logit model has emerged as the model of choice, and most studies in practice have employed some parametric probability function for the mixture distribution. Though parametric distributions are often simpler to specify and easier to estimate, they require the analyst to impose a number of unnecessarily restrictive assumptions that may not always hold true. Nonparametric mixture distributions free analysts from making any assumptions about the shape or functional form of the distribution, but the greater computational burden imposed by the use of these distributions has forced analysts to make simplifying assumptions that do not fully exploit the benefits offered by the more general framework.

This study developed a computationally tractable method for estimating mixed logit models with nonparametric mixture distributions that can help overcome many of the limitations associated with the use of more traditional mixture distributions. The support of the distribution is specified as a high-dimensional grid over the coefficient space, with equal or unequal intervals between successive points along the same dimension, and the location of each point on the grid and the probability mass at that point are model parameters that need to be estimated. Simulated datasets were used to evaluate the ability of the model framework and estimation routine to recover univariate, bivariate and multivariate parametric mixture distributions, and a case study on travel mode choice behavior was employed to assess the value of the framework over models with other mixture distributions.

Our findings indicate that the use of the framework offers three major benefits. First, the shape of the distribution does not need to be specified prior to estimation, and the distribution can mimic any desired probability distribution function to any arbitrary degree of accuracy. Second, behavioral phenomena such as attribute non-attendance and choice set formation can be uncovered endogenously during estimation as special realizations of the general specification, and explicit constraints that achieve the same do not need to be imposed prior to estimation. And third, multivariate distributions with complex covariance structures over high-dimensional coefficient spaces can be estimated more feasibly than would be possible using multivariate parametric distributions.

There are four immediate directions in which future research can build on the work presented here. First, there is considerable room for improvement in the performance of the estimation algorithm. Future research should examine how both the computational efficiency and the statistical robustness of the estimation routine may be improved upon, by leveraging advances in computational hardware and optimization methods. Second, the model framework requires the analyst to specify the number of mass points prior to model estimation. The appropriate number of mass points along any one dimension is subsequently determined by comparing models with differing numbers of mass points along each dimension in terms of both statistical measures of fit and behavioral interpretation. Future research should explore how infinite mixture models, such as Dirichlet process mixture models (see, for example, Burda et al., 2008), that do not require the analyst to determine the appropriate number of mixture components by assuming infinitely many components, may be adapted to overcome this last limitation. Third, due to the nonparametric specification for the mixture distribution, the framework requires the estimation of an exponentially large number of parameters. Future research could examine ways to specify the location of mass points in high-dimensional parameter spaces in more parsimonious ways, for example through the use of seminonparametric distributions such as the one employed by Train (2016). Alternatively, restricted versions of LCCMs with unstructured supports could be used, where only a subset of parameters is allowed to vary across classes. And fourth, we examined the value of the proposed mixing distribution over logit kernels. It'd be interesting to see how findings translate across other discrete choice kernels, such as probit.

More broadly though, as stated by similar calls in the literature (e.g. Wedel et al. 1999 and Ben-Akiva et al., 2012), there is a need for future research to shed greater light on the process underlying differences in taste coefficients. Some studies have related taste heterogeneity to deeper differences in sociological, psychological and biological constructs, such as attitudes, values, perceptions, normative beliefs, affects, lifestyles, etc. (e.g. McFadden, 1986; Ben-Akiva et al., 2002). Others have argued that taste heterogeneity may at times be confounded with heterogeneity along other dimensions of choice behavior, such as decision rules, consideration sets, error structures, etc. (e.g. Swait and Bernardino, 2000). As a collective, the literature on discrete choice analysis has devoted considerable attention towards addressing how best to incorporate random taste heterogeneity, but there is no objectively best way. As analysts, we have the ability to estimate a wide variety of parametric and nonparametric distributions. The appropriate distribution will depend upon the empirical context. There will be cases where the parsimony offered by parametric distributions should be preferred, and there will be cases where the flexibility offered by nonparametric distributions should be preferred.

**Acknowledgements**

The authors wish to thank Kenneth Train, for delivering the lecture that inspired this study; Joffre Swait and Anthony Marley, for their feedback that helped shape the presentation of ideas; John Rose, for pointers to appropriate bodies of relevant literature; the San Francisco County Transportation Authority (SF CTA), for permitting use of their version of the Bay Area Travel Survey (BATS) 2000 dataset; and Chandra Bhat and three anonymous reviewers for their constructive comments. Any mistakes are the fault of the authors alone.

**Appendix A: Monte Carlo Experiments**

Following the methodology proposed by Williams and Ortúzar (1982) and the approach outlined by Raveau et al. (2010), the travel times and costs incurred by different travel modes are synthesized using distributions that are reflective of values that would be observed in empirical data. The model coefficients are specified such that they satisfy three conditions. First, the marginal rates of substitution between the explanatory variables are consistent with values observed by studies in the literature to ensure that the experiment design is as realistic as possible. Second, the part-worth utilities of each of the explanatory variables, as represented by the product between that variable and the corresponding coefficient, are comparable in terms of magnitude. If this is not the case, one of the attributes could potentially dominate the utility function, and it may be hard to empirically isolate the effect of other variables. And third, the scale of the model is set such that the error rate for the data is roughly 25%, i.e. one in four simulated decision-makers change their choice because of the stochastic component, thereby ensuring that the decision-making process is neither completely deterministic nor completely stochastic.

Tables 1 and 2 enumerate the variable and coefficient distributions used by the first experiment, respectively. Tables 3 and 4 enumerate the corresponding distributions used by the second experiment. Table 5 enumerates the variable distributions used by the third experiment.

**Table 1:** Distributions for travel mode level-of-service attributes for experiment 1

| Variable | Notation | Units | Distribution |
|---|---|---|---|
| Travel time - Walk | $tt_{walk}$ | Minutes | $U(1.5tt_{car}, 2.5tt_{car})$ |
| Travel time - Bike | $tt_{bike}$ | Minutes | $U(tt_{car}, 1.5tt_{car})$ |
| Travel time - Car | $tt_{car}$ | Minutes | $U(10, 60)$ |
| Travel time - Transit | $tt_{transit}$ | Minutes | $U(tt_{car}, 2.5tt_{car})$ |
| Cost - Walk | $cost_{walk}$ | $ | 0 |
| Cost - Bike | $cost_{bike}$ | $ | 0 |
| Cost - Car | $cost_{car}$ | $ | $U(0, 20)$ |
| Cost - Transit | $cost_{transit}$ | $ | $U(0, 4)$ |

$U(a, b)$ denotes a uniform distribution over the range $(a, b)$

**Table 2:** Coefficient values for experiment 1

| Coefficient | Notation | Units | Value / Distribution |
|---|---|---|---|
| Constant - Walk | $ASC_{walk}$ | Utils | 0.00 |
| Constant - Bike | $ASC_{bike}$ | Utils | -3.50 |
| Constant - Car | $ASC_{car}$ | Utils | 2.50 |
| Constant - Transit | $ASC_{transit}$ | Utils | 0.50 |
| Travel time | | | |
|   Normal | $\beta_{tt}$ | Utils/minute | $N(-0.9, 0.09)$ |
|   Lognormal | $\beta_{tt}$ | Utils/minute | $LN(-0.5, 0.03)$ |
|   Mixture of normals | $\beta_{tt}$ | Utils/minute | $0.7N(-0.45, 0.02)$ $+0.3N(-1.50, 0.09)$ |
| Travel cost | $\beta_{cost}$ | Utils/$ | -1.80 |

$N(a, b^2)$ denotes a normal distribution with mean a and standard deviation b
$LN(a, b^2)$ denotes a lognormal distribution with location a and scale b

**Table 3:** Distributions for travel mode level-of-service attributes for experiment 2

| Variable | Notation | Units | Distribution |
|---|---|---|---|
| Car | | | |
|   In-vehicle travel time | $ivtt_{car}$ | Minutes | $U(10, 50)$ |
|   Out-of-vehicle travel time | $ovtt_{car}$ | Minutes | $U(0, 10)$ |
|   Speed | $v_{car}$ | Mph | $LN(2.05, 0.63)$ |
|   Distance | $s_{car}$ | Miles | $v_{car} tt_{car}/60$ |
|   Cost | $cost_{car}$ | $ | $U(0, 5) + 0.6 s_{car}$ |
| Walk | | | |
|   Speed | $v_{walk}$ | Mph | $LN(0.28, 0.43)$ |
|   In-vehicle travel time | $ivtt_{walk}$ | Minutes | 0 |
|   Out-of-vehicle travel time | $ovtt_{walk}$ | Minutes | $60(s_{car}/v_{walk})$ |
|   Cost | $cost_{walk}$ | $ | 0 |
| Walk | | | |
|   Speed | $v_{bike}$ | Mph | $LN(1.38, 0.38)$ |
|   In-vehicle travel time | $ivtt_{bike}$ | Minutes | 0 |
|   Out-of-vehicle travel time | $ovtt_{bike}$ | Minutes | $60(s_{car}/v_{bike})$ |
|   Cost | $cost_{bike}$ | $ | 0 |
| Transit | | | |
|   In-vehicle travel time | $ivtt_{transit}$ | Minutes | $U(0.8 tt_{car}, 1.5 tt_{car})$ |
|   Out-of-vehicle travel time | $ovtt_{transit}$ | Minutes | $U(0, 15) + U(0, 15)$ |
|   Cost | $cost_{transit}$ | $ | $U(0, 4)$ |

$U(a, b)$ denotes a uniform distribution over the range $(a, b)$
$LN(a, b)$ denotes a lognormal distribution with location $a$ and scale $b$

**Table 4:** Coefficient values for experiment 2

| Coefficient | Notation | Units | Value / Distribution |
|---|---|---|---|
| Constant - Walk | $ASC_{walk}$ | Utils | 0.00 |
| Constant - Bike | $ASC_{bike}$ | Utils | -3.50 |
| Constant - Car | $ASC_{car}$ | Utils | -1.50 |
| Constant - Transit | $ASC_{transit}$ | Utils | -2.00 |
| In-vehicle travel time | $\begin{bmatrix} \beta_{ivtt} \\ \beta_{ovtt} \end{bmatrix}$ | Utils/hour | $N\left(\begin{bmatrix} -18.00 \\ -54.00 \end{bmatrix}, \begin{bmatrix} 16.20 & 6.48 \\ 6.48 & 32.4 \end{bmatrix}\right)$ |
| Out-of-vehicle travel time | | | |
| Travel cost | $\beta_{cost}$ | Utils/$ | -1.80 |

$N(\mu, \Sigma)$ denotes a multivariate normal distribution with mean $\mu$ and covariance $\Sigma$

**Table 5:** Distributions for alternative-specific attributes for experiment 3

| Variable | Levels | Units | Distribution |
|---|---|---|---|
| Purchase price | continuous | $ 1,000 | U(1.5,8) |
| Operating cost per 100 km | continuous | $ 10 | U(0.5,4) |
| Powertrain technology | Internal combustion engine; hybrid; electric | - | Categorical(1/3 ,1/3 ,1/3) |
| Premium automotive brand | Yes; no | - | Bernoulli(0.4) |

U(a, b) denotes a uniform distribution over the range (a, b)

The five-dimensional parameter distribution corresponding to the third experiment is assumed to be a mixture of three multivariate truncated normal distributions with the following parameters:

$$\boldsymbol{\beta} \sim \sum_{k=1}^{3} w_k v_k$$

$$\{w_k\}_{k=1}^{3} = (0.3, 0.4, 0.3)^T$$

$$v_k \sim N_A(\boldsymbol{\mu_k}, \mathbf{D_k \Omega D_k}), \mathbf{D_k} = \sqrt{\text{diag}(\sigma_k^2 \mathbf{I})}$$

$$\boldsymbol{\mu_1} = (0.00, -0.14, -0.87, -1.30, 1.60)^T$$

$$\boldsymbol{\mu_2} = (-0.94, -1.01, 1.30, 0.07, -1.81)^T$$

$$\boldsymbol{\mu_3} = (-1.81, -1.74, 0.14, 1.45, 0.22)^T$$

$$\boldsymbol{\sigma_1} = (0.20, 0.25, 0.23, 0.35, 0.25)^T$$

$$\boldsymbol{\sigma_2} = (0.15, 0.10, 0.25, 0.35, 0.30)^T$$

$$\boldsymbol{\sigma_3} = (0.40, 0.30, 0.20, 0.30, 0.40)^T$$

$$\boldsymbol{\Omega} = \begin{pmatrix} 1.0 & 0.5 & 0.3 & 0.3 & -0.5 \\ 0.5 & 1.0 & 0.6 & 0.6 & -0.2 \\ 0.3 & 0.6 & 1.0 & 0.3 & -0.4 \\ 0.3 & 0.6 & 0.3 & 1.0 & 0.0 \\ -0.5 & -0.2 & -0.4 & 0.0 & 1.0 \end{pmatrix}$$

$$\mathbf{A} = \left((-\infty, 0), (-\infty, 0), (-\infty, \infty), (-\infty, \infty), (-\infty, \infty)\right)^T$$

**Appendix B: Hierarchical Bayesian logit model**

The hierarchical Bayesian logit model can be derived within the framework of Random Utility Theory (see e.g. Train 2009). For the scope of this study, we assume a linear-in-parameters utility specification with $u_{ntj} = asc_j + \boldsymbol{\beta}_n^T \mathbf{x}_{ntj} + \varepsilon_{ntj}$, where $\boldsymbol{\beta}_n$ is a vector of correlated taste parameters and $\mathbf{x}_{ntj}$ is a vector of covariates; $asc_j$ denotes the alternative-specific constant corresponding to alternative j. For identification, we set $asc_1 = 0$; in the case of unlabelled choice data, $asc_j$ is omitted from the model specification. Otherwise, we let $asc_j \sim N(0, 5^2)$. The assumption $\varepsilon_{ntj} \sim \text{Gumbel}(0, \frac{\pi^2}{6})$ leads to the logit model, which allows us to represent the probability of individual n choosing alternative j on occasion t by

$$P(y_{ntj} = 1 | \mathbf{x}_{nt}; \boldsymbol{\beta}_n, asc_j) = \frac{\exp(asc_j + \boldsymbol{\beta}_n^T \mathbf{x}_{ntj})}{\sum_{j' \in C_{nt}} \exp(asc_j + \boldsymbol{\beta}_n^T \mathbf{x}_{ntj'})}. \quad (1)$$

Using (2), we obtain the probability simplex $\{P(y_{ntj} = 1 | \mathbf{x}_{nt}; \boldsymbol{\beta}_n, asc_j)\}_{j' \in C_{nt}}$, which allows us to specify a categorical distribution, from which $\{y_{ntj}\}_{j' \in C_{nt}}$ is drawn. We have

$$\{y_{ntj}\}_{j' \in C_{nt}} \sim \text{Categorical}\left(\{P(y_{ntj} = 1 | \mathbf{x}_{nt}; \boldsymbol{\beta}_n, asc_j)\}_{j' \in C_{nt}}\right). \quad (2)$$

The individual taste parameters $\boldsymbol{\beta}_n$ are realizations from a multivariate normal distribution defined through mean vector $\boldsymbol{\mu}$ and covariance matrix $\boldsymbol{\Sigma}$ such that $\boldsymbol{\beta}_n \sim N(\boldsymbol{\mu}, \boldsymbol{\Sigma})$. We define a non-informative prior on $\boldsymbol{\mu}$ with $\boldsymbol{\mu} \sim N(0, 5^2)$, but for numerical reasons, we do not directly estimate the covariance matrix $\boldsymbol{\Sigma}$, but rather the scale vector $\boldsymbol{\sigma}$ as well as the Cholesky factor of the correlation matrix $\boldsymbol{\Omega}$. To this end, we exploit the relationships $\boldsymbol{\Sigma} = \mathbf{D}\boldsymbol{\Omega}\mathbf{D}$ with $\mathbf{D} = \text{diag}(\boldsymbol{\sigma})$ and $\boldsymbol{\Omega} = \mathbf{LL}'$. We define priors on $\boldsymbol{\sigma}$ and $\mathbf{L}$: Each element in $\boldsymbol{\sigma}$ is specified to be drawn from the positive half-Cauchy distribution such that $\sigma \sim C^+(0, 2.5)$. A suitable prior for $\mathbf{L}$ is the LKJ Cholesky distribution (Lewandowski et al., 2009), whose density is given by

$$\text{LKJ\_cholesky}(\mathbf{L}|\eta) \propto |J| \det(\mathbf{LL}')^{\eta-1} = \prod_{k=2}^{K} \mathbf{L}_{kk}^{K-k+2\eta-2}, \quad (3)$$

where $\eta > 0$ is a scale parameter. For a discussion of the properties of the LKJ Cholesky distribution, we refer the reader to Stan Development Team (2017). For the purpose of estimating the HB logit model, we let $\mathbf{L} \sim \text{LKJ\_cholesky}(\eta)$ with $\eta=4$.

To estimate the HB logit model specified above, we employ the Stan software package (Carpenter et al., 2016), which implements the No-U-Turn sampler (Hoffman and Gelman, 2016). Prior to the execution of the sampler, we estimate equivalent mixed logit models with uncorrelated random taste parameters via maximum simulated likelihood methods to obtain starting values for the location and scale parameters of the random taste parameters distribution as well as any other fixed taste parameters. For Monte Carlo experiment III, the sampler is executed with four parallel Markov chains and 10,000 iterations for each chain. The initial 2,500 iterations of each chain are discarded for burn-in. Convergence is assessed by considering the Gelman-Rubin diagnostic values (Gelman and Rubin, 1992) and by visually inspecting the trace plots of individual parameter estimates. In the case of the empirical application (Section 6), the sampler becomes prohibitively slow and convergence cannot be attained.

**Appendix C: Parameter estimates for different model specifications for the case study**

| Variable | Multinomial logit without systematic heterogeneity | | Multinomial logit with systematic heterogeneity | | Mixed logit with univariate normals | | Mixed logit with univariate lognormals | | 2048-class LCCM with equal intervals | | 2048-class LCCM with unequal intervals | |
|---|---|---|---|---|---|---|---|---|---|---|---|---|
| | est. | t-stat | est. | t-stat | est. | t-stat | est. | t-stat | est. | t-stat | est. | t-stat |
| *Car specific effects* | | | | | | | | | | | | |
| Constant | - | - | - | - | - | - | - | - | - | - | - | - |
| Standard deviation | - | - | - | - | 5.070 | 13.97 | 4.380 | 11.86 | - | - | - | - |
| *Public transport specific effects* | | | | | | | | | | | | |
| Constant [a] | -0.303 | -7.04 | -0.585 | -5.76 | 4.580 | 6.88 | 4.200 | 5.45 | - | - | - | - |
| Standard deviation | - | - | - | - | 0.019 | 0.01 | 1.610 | 3.72 | - | - | - | - |
| Number of household cars is less than number of household workers | - | - | 1.045 | 17.23 | 0.645 | 3.19 | 0.719 | 3.80 | 3.846 | 18.96 | 4.718 | 21.38 |
| Male | - | - | - | - | - | - | - | - | - | - | - | - |
| 12 years or younger | - | - | -0.739 | -4.92 | -2.130 | -3.32 | -2.080 | -3.08 | -1.905 | -4.21 | -2.304 | -4.65 |
| 13-16 years | - | - | 0.491 | 3.49 | 2.270 | 3.54 | 2.230 | 3.03 | 2.416 | 6.13 | 2.762 | 6.41 |
| 17-24 years | - | - | - | - | - | - | - | - | - | - | - | - |
| 25-44 years | - | - | 0.142 | 1.41 | 0.108 | 0.20 | -0.004 | -0.01 | 0.026 | 0.09 | -0.138 | -0.42 |
| 45-64 years | - | - | -0.113 | -1.08 | -1.180 | -2.21 | -1.240 | -2.25 | -0.587 | -1.88 | -0.789 | -2.32 |
| 65 years or older | - | - | -0.528 | -2.41 | -2.440 | -1.83 | -2.050 | -2.39 | -1.631 | -3.14 | -1.894 | -3.32 |
| *Walk specific effects* | | | | | | | | | | | | |
| Constant [a] | -1.228 | -33.86 | -1.439 | -13.58 | 3.160 | 3.76 | 2.600 | 2.99 | - | - | - | - |
| Standard deviation | - | - | - | - | 3.520 | 9.16 | 3.440 | 4.93 | - | - | - | - |
| Number of household cars is less than number of household workers | - | - | 0.915 | 11.53 | 0.927 | 3.10 | 0.919 | 3.11 | 6.258 | 21.43 | 7.564 | 25.25 |
| Male | - | - | - | - | - | - | - | - | - | - | - | - |
| 12 years or younger | - | - | 0.276 | 2.40 | -1.360 | -1.84 | -1.040 | -1.44 | 0.271 | 0.69 | 0.154 | 0.38 |
| 13-16 years | - | - | 0.740 | 5.70 | 2.530 | 3.06 | 2.850 | 3.26 | 3.146 | 6.92 | 3.613 | 7.62 |
| 17-24 years | - | - | - | - | - | - | - | - | - | - | - | - |
| 25-44 years | - | - | -0.241 | -1.98 | -0.227 | -0.32 | -0.600 | -0.88 | -1.058 | -2.71 | -1.322 | -3.27 |
| 45-64 years | - | - | -0.794 | -5.91 | -3.190 | -4.09 | -3.520 | 4.48 | -2.726 | -6.52 | -3.195 | -7.38 |
| 65 years or older | - | - | -1.369 | -3.67 | -7.500 | -4.51 | -7.130 | -5.13 | -5.811 | -6.65 | -4.890 | -5.86 |

| Variable | Multinomial logit without systematic heterogeneity | | Multinomial logit with systematic heterogeneity | | Mixed logit with univariate normals | | Mixed logit with univariate lognormals | | 2048-class LCCM with equal intervals | | 2048-class LCCM with unequal intervals | |
|---|---|---|---|---|---|---|---|---|---|---|---|---|
| | est. | t-stat | est. | t-stat | est. | t-stat | est. | t-stat | est. | t-stat | est. | t-stat |
| *Bike specific effects* | | | | | | | | | | | | |
| Constant [a] | -3.438 | -60.31 | -4.222 | -23.36 | -13.100 | -4.83 | -10.800 | -6.72 | - | - | - | - |
| Standard deviation | - | - | - | - | 7.690 | 5.69 | 6.780 | 8.70 | - | - | - | - |
| Number of household cars is less than number of household workers | - | - | 1.045 | 9.15 | 0.318 | 0.58 | 0.303 | 0.49 | 2.217 | 6.88 | 3.122 | 8.49 |
| Male | - | - | 1.115 | 9.85 | 5.050 | 4.55 | 5.910 | 8.05 | 2.680 | 10.49 | 2.816 | 10.41 |
| 12 years or younger | - | - | -0.251 | -1.27 | -3.590 | -2.18 | -2.680 | -2.10 | -0.962 | -2.35 | -1.023 | -2.39 |
| 13-16 years | - | - | 0.163 | 0.72 | -1.140 | -0.69 | 0.788 | 0.59 | 0.139 | 0.29 | 0.387 | 0.76 |
| 17-24 years | - | - | - | - | - | - | - | - | - | - | - | - |
| 25-44 years | - | - | 0.251 | 1.44 | 1.030 | 0.60 | 2.870 | 2.43 | 0.446 | 1.15 | 0.455 | 1.11 |
| 45-64 years | - | - | -0.296 | -1.52 | -2.730 | -1.26 | -1.660 | -1.40 | -1.394 | -3.00 | -1.386 | -2.86 |
| 65 years or older | - | - | -0.801 | -1.32 | -3.460 | -1.59 | -4.420 | -3.32 | -2.179 | -2.39 | -1.889 | -2.04 |
| *In-vehicle time (h)* [b] | | | | | | | | | | | | |
| Mean (location) | -1.080 | -73.39 | -1.100 | -73.16 | -5.100 | -14.33 | 1.240 | 16.66 | - | - | - | - |
| Standard deviation (scale) | - | - | - | - | 3.150 | 11.26 | 0.977 | 37.20 | - | - | - | - |
| *Walking time (h)* [b] | | | | | | | | | | | | |
| Mean (location) | -0.641 | -98.85 | -0.587 | -90.31 | -8.510 | -14.89 | 2.020 | 28.47 | - | - | - | - |
| Standard deviation (scale) | - | - | - | - | 4.030 | 14.22 | 0.707 | 36.13 | - | - | - | - |
| *Bicycling time (h)* [b] | | | | | | | | | | | | |
| Mean (location) | -1.254 | -34.33 | -1.331 | -36.09 | -20.300 | -7.77 | 3.820 | 48.47 | - | - | - | - |
| Standard deviation (scale) | - | - | - | - | 8.560 | 7.75 | 1.060 | 35.14 | - | - | - | - |
| *Waiting time (h)* [b] | | | | | | | | | | | | |
| Mean (location) | -0.683 | -13.23 | -0.580 | -11.03 | -3.700 | -7.24 | 0.229 | 0.66 | - | - | - | - |
| Standard deviation (scale) | - | - | - | - | 3.320 | 6.23 | 1.540 | 11.34 | - | - | - | - |
| *Travel cost ($)* | | | | | | | | | | | | |
| Baseline effect | -0.120 | -25.13 | -0.139 | -21.59 | -0.369 | -5.07 | -0.350 | -5.52 | -0.305 | -16.26 | -0.351 | -17.22 |
| Interaction with household income ($100,000s) | - | - | 0.002 | 4.24 | 0.001 | 0.18 | -0.002 | -0.49 | 0.001 | 0.62 | 0.001 | 0.72 |

[a] For the LCCMs, the alternative-specific constants are assumed to be randomly distributed; estimates for the corresponding (finite) mixture distributions are reported in Table 5 in Section 6

[b] For the mixed logit with univariate normals, we report the mean and standard deviation of the corresponding normal random taste coefficient; for the mixed logit with univariate lognormals, we report the location and scale parameters of the negative of the corresponding lognormal random taste coefficient; and for the LCCMS, we refer the reader to Table 5 in Section 6